\documentclass[preprint]{aastex}
\usepackage{mkfig}

\begin{document}

\title{\bf Analysis of H$_{2}$ Emission from Mira~B in UV Spectra
  from HST\altaffilmark{1}}

\author{Brian E. Wood,\altaffilmark{2} Margarita Karovska,\altaffilmark{3}
  and John C. Raymond\altaffilmark{3}}

\altaffiltext{1}{Based on observations with the NASA/ESA Hubble Space
  Telescope, obtained at the Space Telescope Science Institute, which is
  operated by the Association of Universities for Research in Astronomy,
  Inc., under NASA contract NAS5-26555.}
\altaffiltext{2}{JILA, University of Colorado and NIST, Boulder, CO
  80309-0440.}
\altaffiltext{3}{Smithsonian Astrophysical Observatory, 60 Garden St., 
  Cambridge, MA 02138.}

\begin{abstract}

     We analyze Ly$\alpha$ fluoresced H$_{2}$ lines observed in the UV
spectrum of Mira~B.  We identify 13 different sequences fluoresced by 13
different H$_{2}$ transitions within the Ly$\alpha$ line.  The observed
H$_{2}$ line ratios within these sequences imply significant line opacity,
so we use a Monte Carlo radiative transfer code to model the line ratios,
correcting for opacity effects.  We find the observed line ratios can best
be reproduced by assuming that the H$_{2}$ is fluoresced in a layer between
the observer and Mira~B with a temperature and column density of $T=3600$~K
and $\log {\rm N(H_{2})}=17.3$, respectively.  The strengths of H$_{2}$
absorption features within the Ly$\alpha$ line are roughly consistent with
this temperature and column.  We use the total flux fluoresced within the
13 sequences to infer the Ly$\alpha$ profile seen by the H$_{2}$.  In order
to explain differences between the shape of this and the observed profile,
we have to assume that the observed profile suffers additional interstellar
(or circumstellar) H~I Ly$\alpha$ absorption with a column density of about
$\log {\rm N(H~I)}=20.35$.  We also have to assume that the observed
profile is about a factor of 2.5 lower in flux than the profile seen by the
H$_{2}$, and a couple possible explanations for this behavior are presented.
Several lines of evidence lead us to tentatively attribute the fluoresced
emission to H$_{2}$ that is heated in a photodissociation front within
Mira~A's wind a few AU from Mira~B, although it is possible that
interaction between the winds of Mira A and B may also play a role in
heating the H$_{2}$.  We estimate a Mira~B mass loss rate of
$\dot{M}=5\times 10^{-13}$ M$_{\odot}$ yr$^{-1}$ and a terminal velocity of
$V_{\infty}=250$ km~s$^{-1}$, based on wind absorption features in the
Mg~II h \& k lines.  We note, however, that the wind is variable and IUE
Mg~II spectra suggest significantly higher mass loss rates during the IUE
era.

\end{abstract}


\section{INTRODUCTION}

     Mira (o Cet, HD~14386) is one of the most well studied variable stars
in the sky, representing the prototype for the Mira class of pulsating
variables.  Optical spectra of Mira at its minimum brightness reveal the
presence of a hot companion star, Mira~B, which is not easily resolvable
from the ground \citep{ahj26,yy77}.  Nevertheless, both
speckle interferometric techniques applied to ground-based observations
and observations from the {\em Hubble Space Telescope} (HST) have proven
effective at resolving the 2 members of the Mira system, and HST
observations from 1995 show the companion $0.578^{\prime\prime}$ from
the primary at a position angle of $108.3^{\circ}$ \citep{mk91,mk97,mk93}.

     The {\em Hipparcos} distance to the Mira system is $128\pm 18$~pc
\citep{macp97}.  This is a significant increase from a previous
parallax estimate of $d=77$~pc \citep{lfj52}.  The {\em Hipparcos}
distances for Miras (including Mira itself) have been used to define a
new period-luminosity relation for Miras that leads to a distance
measurement to the Large Magellanic Cloud consistent with the accepted
value \citep{fvl97,paw00}.  This provides support for the {\em Hipparcos}
distance to Mira and the other variables in its class.

     Although the optical spectrum of Mira~B is difficult to separate
from that of Mira~A even at Mira~A minimum, \citet{ahj26} estimated a
spectral type of B8 for Mira~B.  The high temperature but low overall
luminosity of Mira~B have led to the assumption that Mira~B is probably a
white dwarf.  However, this conclusion is complicated by the fact that
Mira~B is accreting material from Mira~A's massive cool wind, and the
accretion process clearly affects the appearance of Mira~B's optical
spectrum, resulting in variability on timescales ranging from minutes to
years, with some suggestion of a 14 year periodicity \citep{bw72,yy77}.

     The existence of accretion onto Mira~B makes Mira rather unique in
being a wind accretion system in which one can actually spatially resolve
the accretor from the star whose wind is feeding the accretion.
However, the accretion makes it difficult to be certain whether the
continuum emission we see is from Mira~B or from the
accretion onto the star.  \citet{mj84} argue that
the dearth of X-rays from Mira~B implies that Mira~B cannot be a compact
object like a white dwarf and must instead be a faint main sequence star,
in which case the optical and UV emission is entirely from the accretion
rather than the star itself \citep[see also][]{mk96}.

     The clearest indicators of accretion onto Mira~B are provided by UV
spectra from the {\em International Ultraviolet Explorer} (IUE).  Spectra
taken with IUE show lines such as C~IV $\lambda$1550 and
Si~III] $\lambda$1892, whose broad widths and high temperatures of
formation are most naturally explained by their formation within a hot,
rapidly rotating accretion disk \citep{dr85}.  Mira~B was
observed numerous times during the 18 year lifespan of IUE, the archive
consisting of a total of 94 usable IUE spectra, including both low and
high resolution spectra of both the long and short wavelength regions
covered by the IUE spectrographs.  These data show that the UV continuum
and emission line fluxes vary by a factor of two or so within the data set
\citep{dr85}.  Ultraviolet spectra taken in 1995 with the
Faint Object Camera (FOC) instrument on HST showed fluxes near the low end
of the range observed by IUE, but still consistent with the behavior seen
within the older IUE data set \citep{mk97}.

     However, when the Space Telescope Imaging Spectrometer (STIS)
instrument on HST observed Mira~B on 1999 August 2, the UV spectrum was
dramatically different \citep[][hereafter Paper 1]{bew01}.
Continuum fluxes in these most recent observations
are uniformly more than 10 times lower than ever observed by IUE or HST/FOC.
The STIS observations cover the spectral ranges $1140-1735$~\AA\ and
$2303-3111$~\AA, so this tremendous drop in continuum flux presumably
extends into the optical regime above 3000~\AA.  The temperature of the
gas responsible for the continuum emission apparently did not change, given
that the shape of the continuum did not change.  Thus, the tremendous
change in flux is very hard to explain if the emission is from a stellar
photosphere, but perhaps not so hard to explain if the emission is from the
accreting material itself, in which case a large drop in accretion rate
could account for a drop in flux that would not necessarily be
accompanied by a temperature change.

     The continuum is not the only aspect of the UV spectrum to show
remarkable differences between the IUE era and the HST/STIS observations.
The UV emission lines also varied.  Many lines showed flux decreases
similar to that of the continuum, including the Mg~II h \& k lines near
2800~\AA.  The factor of $\sim 100$ decrease in C~IV $\lambda$1550 flux was
even more extreme.  The P Cygni-like profiles of the Mg~II lines suggest
that accretion onto Mira~B drives a warm, fast outflow.  Comparing the
Mg~II profiles observed by IUE and STIS suggests that the mass loss rate
from Mira~B was significantly lower during the STIS observations than any
time IUE observed these lines.  This is consistent with the idea that the
accretion rate must have dropped substantially, thereby leading to less
mass outflow from Mira~B (see Paper 1).

     However, the most dramatic change in the appearance of the far-UV (FUV)
region of the STIS spectra was not in the continuum or high temperature
line emission but in the appearance of a very large number of narrow
H$_{2}$ lines, which in fact dominate the FUV spectrum observed by STIS
despite not being detected at all in any of the IUE spectra.  In Paper 1,
we announced the discovery of these lines in Mira~B's spectrum, and we
noted that all are Lyman band H$_{2}$ lines fluoresced by the H~I
Ly$\alpha$ line at 1216~\AA.  The Ly$\alpha$ fluorescence mechanism for
exciting UV H$_{2}$ lines has been observed for many astrophysical objects.
The mechanism was first described by \citet{cj77,cj78} to explain
H$_{2}$ lines in the solar spectrum.  Molecular hydrogen lines excited by
Ly$\alpha$ have subsequently been observed in red giant stars
\citep{adm98,adm99}, T Tauri stars \citep{ab81,jav00,dra02a,gjh02}, and
Herbig-Haro objects \citep{rds83,sc95}.

     The first astrophysical detections of Ly$\alpha$ fluoresced H$_{2}$
(excluding the Sun) were from IUE data, but the low sensitivity
of IUE meant that only a few of the strongest H$_{2}$ lines could be
detected, and even these could generally only be detected with low
resolution gratings \citep[e.g.,][]{ab81}.  The Goddard High
Resolution Spectrograph (GHRS) instrument that preceded STIS on board HST
equals STIS in sensitivity and could easily detect the H$_{2}$ lines at
moderate resolution.  However, GHRS only provided limited wavelength
coverage for each exposure, which effectively limited the number of H$_{2}$
lines that could be studied with GHRS \citep[e.g.,][]{sc95,dra02a}.
With STIS, however, one can observe the entire
$1140-1735$~\AA\ region in a single exposure.  Thus, STIS promises to
dramatically increase our understanding of the H$_{2}$ fluorescence process
by allowing detection of many, many more lines than either IUE or GHRS.  In
this paper, we perform a detailed analysis of the H$_{2}$ lines from Mira~B
in order to try to determine where they are coming from, and to see if the
lines can shed light on the nature of the accretion process onto Mira~B.

\section{ANALYSIS}

\subsection{H$_{2}$ Line Identification and Measurement}

     Figure~1 shows the spectral region between 1230~\AA\ and 1650~\AA\
containing all of the H$_{2}$ lines detected in the STIS E140M spectrum,
which in the figure are fitted with Gaussians.  All of the lines can be
identified as members of various Lyman band sequences fluoresced by
the H~I Ly$\alpha$ line.  The Ly$\alpha$ emission from Mira~B excites H$_{2}$
from various rovibrational states in the ground electronic state
($X^{1}\Sigma_{g}^{+}$) to the excited Lyman band electronic state
($B^{1}\Sigma_{u}^{+}$).  From there the H$_{2}$ molecules radiatively
deexcite back to the ground electronic state, resulting in the rich H$_{2}$
line spectrum observed by HST.  We use the \citet{ha93} listing of
Lyman band H$_{2}$ rest wavelengths and transition probabilities to identify
the lines.  To conclusively identify a fluorescence sequence, we require
that at least two lines of the sequence be detected with relative strengths
roughly consistent with the transition probabilities.  In this manner, we
identify 13 fluorescence sequences with 13 different H$_{2}$ transitions
within the H~I Ly$\alpha$ line.  These sequences successfully account for
all of the clearly detected narrow lines in the Mira~B E140M spectrum.

     In Table~1, we provide a complete listing of all 103 positively
detected H$_{2}$ lines, which are grouped into fluorescence sequences.
For H$_{2}$ Lyman band transitions, the dipole selection rules require
$\Delta J\equiv J''-J'=\pm 1$, where $J'$ and $J''$ are the upper and lower
rotational quantum numbers, respectively.  The line ID's in Table~1 use the
standard $v'-v''~X(J'')$ notation, where $v'$ and $v''$ are the upper and
lower vibrational quantum numbers, respectively, and $X=P$ for $\Delta J=+1$
or $X=R$ for $\Delta J=-1$.

     Table~2 lists information on the fluorescing transitions within the
Ly$\alpha$ lines, including rest wavelengths ($\lambda_{rest}$),
absorption strengths (f), and the energies ($E_{low}$) and statistical
weights ($g_{low}$) of the lower levels of the transitions.  Also listed
is the total flux fluoresced within each sequence ($F_{obs}$), in units of
$10^{-13}$ ergs cm$^{-2}$ s$^{-1}$ (see \S 2.2).  The dissociation fraction
per excitation \citep[from][]{ha00} is listed for each
sequence ($f_{dis}^{\prime}$), as is a total dissociation fraction based on
our best model of the H$_{2}$ line ratios ($f_{dis}$; see \S 2.2).

     We measure fluxes for all the lines using Gaussian fits to the lines,
as shown in Figure~1.  When we fit the lines simultaneously, forcing all
lines to have the same centroid velocity and line width, we find an average
centroid of $56.9\pm 0.2$ km~s$^{-1}$, consistent with the central velocity
measured for Mira's circumstellar shell of about 56 km~s$^{-1}$
\citep{pfb88,pp90,ej00}.  The measured
line fluxes listed in Table~1 are from a fit in which the centroids were
allowed to vary but the line widths are still forced to be the same.  The
H$_{2}$ line width measured in this fit is $FWHM=19.7\pm 0.4$ km~s$^{-1}$,
after correction for the E140M line spread function from \citet{kcs99}.

     We then tried a different fit to the data to see if the line ratios
within the 13 fluorescence sequences are consistent with the transition
branching ratios from \citet{ha93}.  We simultaneously fitted all
of the lines, forcing all to have the same centroid velocity and line
width, but we also forced the line flux ratios to be consistent with the
transition probabilities.  What this means is that besides the average line
centroid and line width, the only other free parameters of the fit are 13
flux normalization factors for the 13 sequences of lines.  The result is
shown in Figure~2, where a smoothed representation of the entire E140M
spectrum is compared with the best fit.  The fit is very poor, with
the discrepancy between the observation and model clearly being
wavelength-dependent.  In particular, the shorter wavelength lines are
observed to have lower fluxes relative to the longer wavelength lines than
the transition probabilities suggest.

     We believe that this is due to an opacity effect.  We will show in
\S 2.3 and \S 2.4 that there are actually visible H$_{2}$ absorption
features within the Ly$\alpha$ line where some of the H$_{2}$ sequences are
being pumped.  This means that the H$_{2}$ lines being fluoresced have
significant opacity.  Lower wavelength lines are transitions to lower
vibrational levels than the higher wavelength lines.  The lower vibrational
levels will have higher populations if the levels are thermally populated,
so the lower wavelength lines should have higher opacities, thereby
explaining why they have systematically lower fluxes than the transition
probabilities predict.

\subsection{Radiative Transfer Simulations}

     In order to determine whether opacity effects can in fact explain why
the observed H$_{2}$ line ratios are different from the transition
branching ratios, we perform Monte Carlo radiative transfer simulations,
which we now describe in detail.  The fact that we see the H$_{2}$ in
absorption within Ly$\alpha$ (see \S 2.3 and \S 2.4) suggests that we see
the Ly$\alpha$ emission through an H$_{2}$ layer.  Thus, in
our model we construct a plane-parallel slab of H$_{2}$ with an assumed
H$_{2}$ column density, $N(H_{2})$, and temperature, T.  We assume
Ly$\alpha$ emission is normally incident on one side of the slab, with our
observations being made from the other side of the slab.

     The wavelength dependent opacity of the slab to a given H$_{2}$
transition from a lower level with column density, $N_{pop}$, is (in cgs
units)
\begin{equation}
\tau_{\lambda}=0.02654 f N_{pop} \phi_{\lambda},
\end{equation}
where $f$ is the oscillator absorption strength of the transition and
$\phi_{\lambda}$ is a Voigt line profile function with a width
appropriate for the assumed temperature.  We compute the f-values from
the transition probabilities of \citet{ha93}, using the relation
between the two from \citet{ha89}.  Assuming the H$_{2}$ levels
are thermally populated, the column density of H$_{2}$ at a given level,
$N_{pop}$, depends on both $N(H_{2})$ and T, and can be computed from the
following equation:
\begin{equation}
N_{pop}(J''=J,v''=v)=N({\rm H_{2}})\frac{g_{low}(J,v)\exp
  \left( -1.439E_{low}(J,v)/T \right)}
  {\sum_{J'',v''} g_{low}(J'',v'')\exp
  \left( -1.439E_{low}(J'',v'')/T \right)},
\end{equation}
where $E_{low}$ is the energy of the lower level of the H$_{2}$
transition (in cm$^{-1}$) and $g_{low}$ is the statistical weight of this
level.  The H$_{2}$ energy levels that we assume are from \citet{id84}.
The statistical weights are computed from the equation
\begin{equation}
g_{low}=(2J''+1)(2I+1),
\end{equation}
where I is the nuclear spin, which is 0 for even $J''$ and 1 for odd $J''$
\citep{dcm76}.  Note that Table~2 lists $f$, $E_{low}$, and
$g_{low}$ values for the fluorescing transitions within Ly$\alpha$.

     We model the line ratios of each fluorescence sequence separately.
For the sequence fluoresced by the $1-2$~R(6) transition (see Table~1), for
example, we numerically send Ly$\alpha$ photons across the absorption line
profile ($\phi_{\lambda}$) of the $1-2$~R(6) line.  In the Monte Carlo
routine, the number of optical depths a photon travels before scattering is
the negative natural logarithm of a random number generated between 0 and 1.
When this number is less than the H$_{2}$ optical depth of the slab seen by
the Ly$\alpha$ photon, the photon is absorbed by the H$_{2}$.  The excited
H$_{2}$ molecule then either deexcites or is dissociated.  We use the
dissociation probabilities of \citet*{ha00} to indicate
the fraction of excitations to the Lyman band that result in dissociation
($f_{dis}^{\prime}$), which are listed in Table~2 for each sequence.

    If the molecule radiatively deexcites, the result is an H$_{2}$ emission
photon.  We use the transition probabilities of \citet{ha93} to
indicate how the photons are to be distributed among the various lines in
the sequence (i.e., the $1-x$~R(6) and $1-x$~P(8) sequences for the example
sequence mentioned above).  We randomly determine a new location for the
photon within the line profile ($\phi_{\lambda}$) assuming complete
frequency redistribution to establish the probability distribution.  We
also randomly determine a new direction for the photon assuming complete
angular redistribution.

     We compute the H$_{2}$ opacity for the new photon to escape the slab
based on where it is in the slab, what particular line it is in, where it
is in the line profile, and what direction it is traveling, and then we
repeat the process of determining whether the photon escapes the slab or is
absorbed by an H$_{2}$ molecule again.  Photons that ultimately scatter
back towards the Ly$\alpha$ source are discarded.  Photons that scatter out
the other end of the slab towards the observer are counted.
By numerically sending enough photons through the slab for each of the 13
fluorescence sequences, we can determine what the line ratios will be for
each sequence, for comparison with the observed line fluxes.  We use a
separate fitting routine to determine the best 13 normalization factors to
convert the model photon counts to the observed line fluxes of the 13
fluorescence sequences.  We perform this whole simulation for a range of
assumed T and $N(H_{2})$ values.

     Figure~3 displays $\chi^{2}_{\nu}$ contours showing the quality of
the fit to the data as a function of T and $N(H_{2})$.  The best fit is
for $T=3600$~K and $\log N(H_2)=17.3$ and has $\chi^{2}_{\nu}=4.88$.
The temperature and column density are highly dependent on each other,
however, and Figure~3 shows that there is a long trough in the
$\chi^{2}_{\nu}$ contours, implying that uncertainties in $T$ and $N(H_{2})$
could be large.  Because systematic uncertainties surely dominate this type
of analysis, we do not attempt to quote error bars based on the
$\chi^{2}_{\nu}$ values.  However, in the following sections we will present
two other arguments for these measurements being reasonable and consistent
with the data.  In any case, 3600~K is a plausible temperature for the
H$_{2}$ in the sense that the temperature is not so high that the H$_{2}$
should be collisionally dissociated.

     Figure~4 shows the model spectrum for the best fit from our
simulations, compared with the observed spectrum.  Comparing Figure~4 with
Figure~2 reveals that taking into account opacity effects can indeed remove
the large wavelength dependent discrepancy noted above.  However, there are
still some discrepancies between the model and the data that are more
easily seen in Figure~5, which compares the fluxes of this best model with
the observed fluxes.  There are many data points that lie farther from the
data-model agreement line than the 1 $\sigma$ measurement uncertainties
indicated in the figure.  The model line fluxes are also listed in Table~1
in order to indicate explicitly which lines fit well and which do not.

     In Figure~6, we plot the flux discrepancies between the best model and
the data (in standard deviation units) as a function of wavelength.  Lines
in fluorescence sequences that are excited from low energy levels
($E_{low} < 10,000$ cm$^{-1}$) are shown as open boxes and lines of
fluorescence sequences excited from high energy levels ($E_{low} > 10,000$
cm$^{-1}$) are shown as filled boxes.  (The $E_{low}$ values for all the
sequences are listed in Table~2.)  It is apparent that the low energy
lines are systematically more discrepant than the high energy lines.  It
is tempting to claim that this might be evidence of multiple H$_{2}$
temperature components, since in principle the low energy lines could be
excited at lower temperatures than the high energy lines.  However, there
is no obvious wavelength dependence in the flux discrepancies that would
suggest that the estimated $T$ value is inaccurate for the low energy lines,
and when we perform the analysis described
above considering only the low energy lines we do not obtain significantly
different results --- the large discrepancies for the low energy lines
remain.  Therefore, it is uncertain what is indeed responsible for this
behavior, but perhaps the lower energy levels of H$_{2}$ are not solely
populated by thermal collisions.

     Our best fit to the data, which is shown in Figure~4, provides us with
measurements of the total flux fluoresced by each of the 13 fluorescence
transitions within Ly$\alpha$, $F_{obs}$, which are listed in Table~2.
These total fluoresced fluxes include the fluxes of lines within each
sequence that are undetected either because they are too weak or because
they lie outside the observed spectral range.  We can include these
unobserved fluxes because our fit makes predictions for what the fluxes of
these lines should be.  Our fit also tells us the fraction of fluorescences
that lead ultimately to dissociation of an H$_{2}$ molecule rather than the
emergence of an H$_{2}$ emission photon.  This is recorded in the
dissociation fraction, $f_{dis}$, in Table~2.  The $f_{dis}$ fraction is
naturally larger than the dissociation fraction per excitation,
$f_{dis}^{\prime}$, because the H$_{2}$ opacity sometimes results in
multiple excitations before an H$_{2}$ photon either escapes or
dissociates an H$_{2}$ molecule.

\subsection{Deriving the Ly$\alpha$ Profile Seen by the Fluoresced H$_{2}$}

     The amount of Ly$\alpha$ flux absorbed by a given H$_{2}$
transition depends in part on the amount of H~I Ly$\alpha$ flux overlying
the transition ($f_{\circ}$).  An absorption line profile, $f(\lambda)$,
can be expressed as
\begin{equation}
f(\lambda)=f_{\circ} \exp(-\tau_{\lambda}),
\end{equation}
where the opacity profile, $\tau_{\lambda}$, is given by equation (1).
After accounting for the fraction of H$_{2}$ fluorescences that lead
to dissociation rather than an H$_{2}$ photon, the total emission observed
from a fluorescence sequence ($F_{obs}$) will be equal to the absorbed flux
along a line of sight if the H$_{2}$ that is being fluoresced completely
surrounds the star and the H$_{2}$ emission emerges isotropically.  The
emission will be less than this if the H$_{2}$ only partially surrounds the
star, or if the H$_{2}$ emission is preferentially scattered away from our
line of sight toward Mira~B.  The emission could be greater if the emission
is preferentially scattered {\em toward} our line of sight.  Thus,
\begin{equation}
F_{obs}=\eta \left( 1-f_{dis} \right) \int (f_{\circ}-f(\lambda))d\lambda,
\end{equation}
where $\eta$ is a correction factor meant to account for inaccuracies in
the assumptions of uniform H$_{2}$ coverage and isotropic emission.  If the
emission is assumed isotropic, $\eta$ is then simply a number between 0 and
1 representing the H$_{2}$ coverage fraction.  The $(1-f_{dis})$ factor
corrects for dissociations, and the $f_{dis}$ values can be found in
Table~2.  Combining equations
(4) and (5),
\begin{equation}
\eta f_{\circ} = \frac{F_{obs}}{1-f_{dis}}
  \left( \int 1-\exp(-\tau_{\lambda}) d\lambda \right)^{-1}.
\end{equation}
Table~2 lists $F_{obs}$ values for our 13 fluorescence sequences.  The
opacity profile of the line being fluoresced, $\tau_{\lambda}$, can be
computed from equations (1)--(3), where Table~2 lists the $f$, $E_{low}$,
and $g_{low}$ values for the fluorescing transitions.  Assuming the
$T=3600$~K and $\log N(H_{2})=17.3$ values derived in the previous section,
we can then compute $\eta f_{\circ}$ values from equation (6) for each of
our fluorescence sequences.

     These quantities should map out the Ly$\alpha$ profile that is seen by
the H$_{2}$, when plotted as a function of the wavelength of the
fluorescing transition.  This is what the boxes indicate in Figure~7,
which demonstrates that the $\eta f_{\circ}$ values do indeed map
out a reasonably self-consistent line profile, although there is still
some scatter.  We interpolate between the boxes (and extrapolate beyond
them) to obtain the dashed line in figure.  The H$_{2}$ flux fluoresced by
the 0-2 R(2) line at 1219.09~\AA\ is low based on Figure~7.  This is likely
caused by the blending of this line with the 2-2 R(9) line at 1219.10~\AA,
which results in the 2-2 R(9) line shielding the 0-2 R(2) line from the
full Ly$\alpha$ flux at that wavelength.

     Figure~7 compares the H~I Ly$\alpha$ profile seen by the H$_{2}$
(dashed line) with the Ly$\alpha$ profile that we actually observe in
our STIS data (dotted line).  There appear to be absorption features
within this profile, some of which line up with the H$_{2}$ lines.  We
will return to this issue in the next subsection.  Both the observed
and derived Ly$\alpha$ profiles in Figure~7 are almost entirely redward
of the rest frame of the star, which is the vertical dot-dashed line in the
figure.  As discussed in Paper 1, the Mg~II h \& k lines of Mira~B show a
similar behavior, and the cause is presumably a fast wind from Mira~B that
absorbs the blue side of the line.  {\em The fact that the profile seen by
the H$_{2}$ also has no emission on the blue side means that the wind
absorption happens before the Ly$\alpha$ emission encounters the H$_{2}$,
which is an important clue to the location of the H$_{2}$ being fluoresced.}

     There are many other H$_{2}$ transitions between 1210~\AA\ and
1222~\AA\ that could potentially have been fluoresced, if the lower levels
had been populated and if there was sufficient background flux at the
appropriate wavelength.  Thus, we estimated upper limits for the fluoresced
flux from these transitions and these upper limits are used to compute the
upper limits to $\eta f_{\circ}$ shown in Figure~7.  To be more precise,
for each undetected fluorescence sequence we identify the easiest to detect
line within the sequence by dividing the transition probability of each
line by the flux error of our observed spectrum at the wavelength of that
line --- the line with the largest value is assumed to be the easiest to
detect.  We estimate a conservative upper limit for the flux of this line
and then extrapolate that flux to a total fluoresced flux ($F_{obs}$)
assuming the line ratios within the sequence are those suggested by the
transition probabilities.  This does not take into account the opacity
effects that we see in the detected sequences, but as long as we are
conservative in our initial flux upper limit the final total flux upper
limit should be acceptable.  Using this upper limit for $F_{obs}$, we can
then compute an upper limit for $\eta f_{\circ}$ from equation (6).  We
correct for dissociation assuming $f_{dis}=1.2f_{dis}^{\prime}$, which is
roughly what we see for our detected sequences in Table~2.  In this way,
we compute all the upper limits shown in Figure~7.

     There are two important results from the upper limit calculations.
The first is that there are no upper limits beneath the boxes that would
suggest nonthermal populations (i.e., there are no levels that should have
been thermally populated at $T=3600$~K but are not).  The second is that
there clearly {\em are} transitions that would have been fluoresced by the
blue side of the Ly$\alpha$ line if the flux on that side of the line was
comparable to that on the red side.  This confirms the conclusion stated
above that the blue side of the Ly$\alpha$ line must be absorbed before the
Ly$\alpha$ emission encounters the H$_{2}$.

     The relatively self-consistent, smooth Ly$\alpha$ profile suggested
by the boxes in Figure~7 implies that the $T=3600$~K and $N(H_{2})=17.3$
values derived in the previous section and assumed in the derivation of
the Ly$\alpha$ profile are reasonably accurate measurements.  If these
values were wildly inaccurate or if the H$_{2}$ energy levels were not
thermally populated at all, then the boxes would not line up like they
do in Figure~7.  Figure~8 shows what happens if we arbitrarily assume
temperatures of 2600~K and 4600~K in this analysis instead of 3600~K.
For the low temperature case (Fig.~8a), the scatter in the data points
clearly increases compared to Figure~7.  For the high temperature case
(Fig.~8b), some of the upper limits fall too low.

     We do not expect our derived Ly$\alpha$ profile in Figure~7 (dashed
line) to precisely reproduce the observed profile (dotted line) for two
primary reasons:  1. The Ly$\alpha$ profile could suffer from additional
circumstellar and/or interstellar H~I absorption {\em after} it encounters
the H$_{2}$ region that is fluoresced, making the observed profile lower
than the derived profile.  2. The geometrical correction factor $\eta$
could be different from one.  Since the observed profile is lower than the
derived profile, additional absorption of Ly$\alpha$ (case \#1) is a
distinct possibility.  Conclusive evidence of additional absorption beyond
the H$_{2}$ region is provided by the simple fact that there are two
fluoresced transitions near 1216~\AA\ where there is no observed Ly$\alpha$
flux whatsoever.

     We now try to determine if interstellar or circumstellar absorption
alone can account for the difference between the observed and derived
Ly$\alpha$ profiles in Figure~7.  In order to do this, we first have to
estimate the centroid and temperature of the absorbing material for both
the ISM and circumstellar absorption cases.  The local cloud flow vector of
\citet{rl95} predicts a heliocentric velocity for the ISM
absorption of 18.6 km~s$^{-1}$.  For a line of sight as long as that
towards Mira there will undoubtedly be many ISM absorption components that
will cause the actual centroid of the ISM absorption to differ
significantly from this value, but the value should nevertheless be
accurate enough for our purposes.  For the ISM temperature, we assume a
typical temperature for warm ISM material of 8000~K \citep[e.g.,][]{bew98}.
For circumstellar material, which should be expanding only very
slowly, we assume a heliocentric velocity equal to Mira's estimated radial
velocity of 56 km~s$^{-1}$, and we assume a cold temperature of 500~K.

     Based on these assumptions, in Figure~9a we show the ISM and
circumstellar absorption that results from column densities of
$\log N(H~I)=20.3$, 20.5, and 20.7.  The ISM and circumstellar absorption
profiles are very similar, emphasizing the fact that at column densities
this high the absorption profiles are not very dependent on either the
temperature or centroid velocity.  However, none of the column densities
results in an acceptable fit to the data.  There is simply too much
absorption in the far red wing.  When column densities are raised high
enough to account for it, there is then way too much absorption between 1217
and 1219~\AA.  The only way we can think of rectifying this situation
is to assume that our correction factor, $\eta$, is greater than one.
This could mean that the H$_{2}$ emission is for some reason preferentially
scattered into our line of sight to Mira~B, meaning we should scale the
model Ly$\alpha$ profile downwards.  It could also mean that the
Ly$\alpha$ flux is preferentially scattered {\em out} of our line of sight,
which would mean we should scale our observed Ly$\alpha$ profile upwards.
We will discuss these two possibilities in more detail below (see \S 2.5).
For now, we assume the latter possibility is the case and by trial and
error we find that we can best fit the data when we scale the observed
Ly$\alpha$ profile upwards by a factor of 2.5 (i.e., $\eta=2.5$), in which
case a column density of $\log N(H~I)=20.35$ yields a reasonable fit to the
data, as shown in Figure~9b.

     In the figure, the ISM absorption fits a little better than the
circumstellar absorption.  We attach no importance to this, however,
since the fit to the circumstellar absorption can be improved by slight
adjustments to the flux scaling factor or $\log N(H~I)$ value.  An estimate
of the mass of Mira's circumstellar envelope presented by \citet{pfb88}
from observations of the H~I 21~cm line suggests a
circumstellar column density of only $\log N(H~I)=18.7$, which is much too
low to account for the observed H~I absorption.  We have some concerns that
trying to estimate the amount of circumstellar H~I from the 21~cm emission
could result in an H~I mass that is much too low, since the lifetime of the
21~cm transition ($\sim 10^{7}$ yrs) is so much greater than the crossing
time across the circumstellar envelope ($\sim 10^{4}$ yrs for a typical
expansion velocity of 5 km~s$^{-1}$), meaning that much of the H~I
that is collisionally excited will actually leave the envelope before
emitting a 21~cm photon.  Thus, we do not completely rule out the
possibility that the H~I absorption is circumstellar.

     As for the possibility that the H~I absorption is interstellar, the
column density of $\log N(H~I)=20.35$ is not impossibly high for Mira's
distance of 128~pc.  Mira is likely to be outside the Local Bubble (where
ISM column densities are quite low) based on the Local Bubble maps
estimated by \citet{dms99} from Na~I column density measurements.
One of the stars in their list that is near Mira's location
($l=168^{\circ}$, $b=-58^{\circ}$, d=128~pc) is HD~14613
($l=192^{\circ}$, $b=-67^{\circ}$, d=121~pc), which shows a high Na~I
column of $\log {\rm N(Na~I)}>12.08$, probably corresponding to an H~I
column well above $10^{20}$ cm$^{-2}$ \citep{dms99}.

\subsection{The Evolution of the Ly$\alpha$ Profile During its Journey
  Towards Earth}

     In Figure~10, we present a model of how the Ly$\alpha$ profile evolves
along the line of sight from Mira~B to HST based on the results of the
previous subsection.  The starting point of the
model is the solid line in Figure~10a, which is simply the dashed line
from Figure~7.  As described above, this is the profile seen by H$_{2}$
after Mira~B wind absorption has already absorbed the blue half of the
profile.  The H~I Ly$\alpha$ opacity in the wind figures to be so high that
scattering will erase all memory of the original Ly$\alpha$ profile, but we
can at least obtain a schematic representation for the original profile
by reflecting our derived profile from Figure~7 onto the blue side and then
extrapolating between the peaks.  The result is the dotted line in
Figure~10a.

     Thus, the accretion disk around Mira~B produces a Ly$\alpha$ profile
that looks schematically like the dotted line in Figure~10a.  Mira~B's
wind erases the blue side of the profile, resulting in the solid line
profile in Figure~10a.  The Ly$\alpha$ emission then encounters the
$T=3600$~K, $\log N(H_{2})=17.3$ H$_{2}$, resulting in narrow H$_{2}$
absorption features within the Ly$\alpha$ profile, as shown in Figure~10b.
The Ly$\alpha$ emission then makes the long journey through both Mira's
large circumstellar envelope and the ISM.  This results in
$\log N(H~I)=20.35$ worth of Ly$\alpha$ absorption, the effects of which
are illustrated in Figure~10c.  The resulting profile is smoothed with the
instrumental profile appropriate for our STIS E140M observation, and in
Figure~10d that profile is compared with the observed Ly$\alpha$ profile
multiplied by a factor of 2.5 (see \S2.3).  Our best model of the H$_{2}$
line ratios predicts that some of the Ly$\alpha$ flux that is absorbed by
the H$_{2}$ will naturally be reemitted in the H$_{2}$ lines within
Ly$\alpha$ that are being fluoresced (see, e.g., Fig.~4).  Thus, when
comparing the observed Ly$\alpha$ profile with the model, we first remove
the H$_{2}$ emission component from the Ly$\alpha$ profile by subtracting
the H$_{2}$ emission line fluxes predicted by our model from the observed
profile.  This is only a minor correction, but it does slightly deepen the
apparent H$_{2}$ absorption features within the observed Ly$\alpha$ profile
that were noted above.  Finally, as noted above the observed profile is
multiplied by a factor of 2.5 to match the fluxes of the model profile in
Figure~10d.  Both profiles in Figure~10d were smoothed additionally by a
5-bin boxcar in order to reduce noise in the observed Ly$\alpha$ line.

     The observed Ly$\alpha$ profile does suggest the presence of
absorption features where the model suggests there should be H$_{2}$
absorption (see Fig.~10d), validating our initial assumption that the
fluoresced H$_{2}$ does indeed lie at least in part between us and
Mira~B.  The $T=3600$~K and $\log N(H_{2})=17.3$ measurements based on the
result from Figure~3 slightly overpredict the amount of observed
absorption for most lines.  Based on the H$_{2}$ absorption alone, we would
estimate an H$_{2}$ column density along the observed line of sight to be
more like $\log N(H_{2})\approx 17.1$ (assuming $T=3600$~K naturally).
There could in principle be a difference between the amount of H$_{2}$
along the line of sight we are observing and the average line of sight from
Mira~B.  However, the observed amount of absorption is consistent enough
with that predicted by the model to provide further support for the
$T=3600$~K and $\log N(H_{2})=17.3$ measurements from the line
ratio analysis.

\subsection{The Discrepancy Between the Observed Ly$\alpha$ and H$_{2}$
  Fluxes}

     One aspect of our analysis of the H$_{2}$ lines that has yet to be
explained is this factor of 2.5 that we must multiply the observed
Ly$\alpha$ profile by in order to be able to account for the amount of
H$_{2}$ emission produced by Ly$\alpha$ fluorescence (see Fig.~9).  Why do
we observe weaker Ly$\alpha$ fluxes than the H$_{2}$ must see?  There were
2 basic possibilities suggested above in \S 2.3.  One is that the H$_{2}$
emission is preferentially scattered into our line of sight, and the other
is that the Ly$\alpha$ emission is preferentially scattered {\em out} of
our line of sight.

     One way the first possibility might occur is if there is an optically
thick H$_{2}$ layer of some sort behind both the Ly$\alpha$ source and the
H$_{2}$ layer seen in absorption within Ly$\alpha$.  Such a
background layer could act to scatter all the H$_{2}$ emission back in our
direction.  This would mean that if Mira~B were observed from the opposite
direction, $180^{\circ}$ from the actual line of sight, little if any
H$_{2}$ emission would be observed, whereas in our line of sight we would
see about twice as much H$_{2}$ emission as would be observed if the
emission was isotropic.  The problem with this scenario is that the
backscattering H$_{2}$ layer would have to have a high enough column
density that it should produce more H$_{2}$ flourescence sequences than we
actually observe.  In essence, one would have a situation analogous to
Figure~8b, where the upper limits are becoming inconsistent with the
detections.  There does not seem to be an easy way to preferentially
scatter H$_{2}$ into our line of sight without introducing this problem.

     This is why we prefer the second possibility mentioned above, that
the Ly$\alpha$ emission is preferentially scattered out of our line of
sight.  The most likely scenario for how this could happen has to do with
the possibility that we are observing Mira~B's accretion disk close to
edge-on, much as it is depicted in Figure~11.  The orbit of Mira~B around
Mira~A is poorly known.  \citet{jlp02} have derived a new orbit that is
more consistent with recent observations \citep*{mk93} than an older
estimate from \citet{pb80}.  The \citet{jlp02} orbit suggests an orbital
period of 498 days and an inclination angle of $i=112.0^{\circ}$.  Assuming
that the accretion disk lies in the same plane as Mira~B's orbit, the disk
is then observed only $22.0^{\circ}$ from edge-on.  However, since
observations currently cover only about 15\% of the orbit, the
\citet{jlp02} orbit must still be considered very uncertain at this point.

     Nevertheless, if the accretion disk is close to edge-on there are two
ways that the Ly$\alpha$ emission could be suppressed along our line of
sight.  One is that dust in the disk extends high enough above the plane
of the disk to extinct the observed Ly$\alpha$ emission by a factor of 2.
In this scenario, most of the H$_{2}$ being fluoresced will be above the
polar regions of the disk and will see the Ly$\alpha$ emission without any
extinction, thereby explaining why the H$_{2}$ emission fluxes seem higher
than the observed Ly$\alpha$ fluxes would predict.  A second possibility
relies on the fact that Mira~B's Ly$\alpha$ emission has to pass through a
wind that is extremely opaque, which accounts for the highly redshifted
nature of the emission that ultimately emerges (see Fig.~7).  If H~I
densities in Mira~B's wind are significantly higher closer to the plane of
the disk, this could mean that the Ly$\alpha$ emission preferentially
escapes in less opaque polar directions, explaining why we see less flux
in our supposed edge-on line of sight.  \citet{dra02b} present
evidence that T~Tauri star winds are in fact denser near the
accretion disks.

\subsection{The Location of the Fluoresced H$_{2}$:  The Wind Interaction
  Region?}

     There are basically two possible locations for the observed H$_{2}$
that we consider.  One is that the H$_{2}$ is in the outer regions
of the accretion disk, and the other is that the H$_{2}$ is within the
wind of Mira~A.  Fluoresced H$_{2}$ emission observed from T~Tauri
stars can apparently be either unresolved emission from the accretion disk,
or emission from an outflow, in which the H$_{2}$ may be circumstellar
material entrained in the wind \citep{dra02a,gjh02}.
For T~Tauri itself, the H$_{2}$ emission is clearly extended and is
therefore probably circumstellar \citep{ab81,tmh97}.

     For Mira~B, there are several factors that make the accretion disk
interpretation less attractive.  One is the presence of H$_{2}$ absorption
within the Ly$\alpha$ line, meaning the disk H$_{2}$ would have to exist
above the disk far enough to extend into our line of sight.  Furthermore,
in such a situation one would expect $\eta \ll 1$ since the disk H$_{2}$
would only intercept a small fraction of the total Ly$\alpha$ emission.
The actual value of $\eta$ ($\eta=2.5$ from \S 2.3) is more consistent with
the H$_{2}$ completely surrounding the star, which is difficult to
reconcile with a disk interpretation.  The narrow widths of the H$_{2}$
lines may also be a problem.  As discussed in \S2.5, the disk is observed
only $22.0^{\circ}$ from edge-on based on the \citet{jlp02} orbit, so if the
H$_{2}$ lines were in the disk their widths should be at least as broad as
the rotational broadening provided by disk rotation.  If the mass of Mira~B
is $\sim 0.5$~M$_{\odot}$, the observed width of the H$_{2}$ lines
($FWHM=19.7\pm 0.4$ km~s$^{-1}$) would suggest a distance from the star of
at least 1.1~AU, assuming Keplerian rotation.  The disk would have to be
flared significantly to expose parts of the disk this far out to the
Ly$\alpha$ emission.  One might also expect rotationally broadened line
profiles to be non-Gaussian, but they are in fact Gaussian in shape.

     Thus, we believe that the H$_{2}$ is more likely to be within Mira~A's
wind, which will surround the star (consistent with $\eta=2.5$) and will
be beyond the Mira~B wind absorption region, consistent with our finding
from Figure~7 that the H$_{2}$ fluorescence must come after Mira~B's wind
absorbs the blue half of the Ly$\alpha$ line.  Hydrogen in Mira~A's
massive wind should be largely molecular \citep{pfb88},
but it should be significantly colder than the H$_{2}$ temperature of
$T=3600$~K that we derive.  Thus, the H$_{2}$ in Mira~A's wind must be
heated somehow as it approaches Mira~B.  One possibility is that the
H$_{2}$ is heated by its interaction with the hotter and faster, but less
massive wind of Mira~B, placing the fluoresced H$_{2}$ in the interaction
region between the two winds, as shown schematically in Figure~11.

     In order to ascertain what the wind interaction region of the Mira
binary system might be like, it is necessary to review what is known about
the winds of the 2 stars.  There have been many estimates for Mira~A's mass
loss rate and wind velocity from observations of CO, H~I, and Ca~II in Mira's
circumstellar environment.  The mass loss estimates have generally fallen
in the range $4\times 10^{-8}$---$4\times 10^{-7}$ M$_{\odot}$ yr$^{-1}$,
and wind velocities have generally been quoted as $3-7$ km~s$^{-1}$
\citep{yy78,grk85,pfb88,pp90,grk98,nr01}.

     The accretion rate onto Mira~B that results from Mira~A's wind has
been estimated to be about $8\times 10^{-10}$---$3\times 10^{-9}$
M$_{\odot}$ yr$^{-1}$, which can account for Mira~B's total luminosity of
about 0.2~L$_{\odot}$ \citep{bw72,mj84,dr85}.
The accretion rate provides an upper limit for Mira~B's
mass loss rate.  However, although the existence of Mira~B's wind has been
noted before on the basis of absorption within the H~I Balmer lines
\citep{yy77}, there has been no estimate made for the mass
loss rate.  Thus, we now derive our own estimate.

     Probably the best diagnostics for Mira~B's wind are the Mg~II h \& k
lines at 2800~\AA.  Our 1999 STIS observations include an observation with
the moderate resolution E230M grating of the 2303--3111~\AA\ region
containing Mg~II, and IUE observed the Mg~II lines many times with its high
resolution grating.  In Figure~12, we show the Mg~II k line profile
observed by STIS, and a typical Mg~II k line profile observed by IUE
(LWP 29795 from 1994 December 30), both plotted on a velocity scale
centered in the stellar rest frame.  These spectra were also presented
in Paper 1.  Both Mira~A and Mira~B are in the aperture for the IUE
observations and emission from Mira~A is occasionally seen by IUE
at certain pulsation phases \citep[see, e.g.,][]{bew00},
but for the spectrum in Figure~12b all the emission is from Mira~B.

     The IUE spectrum shows saturated wind absorption between 0 and $-400$
km~s$^{-1}$.  The IUE data set of 28 high resolution spectra containing
Mg~II shows that the Mg~II emission and the wind absorption are clearly
variable, with the extent of the absorption varying between $-200$ and
$-600$ km~s$^{-1}$.  However, the absorption is always very opaque, in
contrast to the STIS Mg~II line, which only shows opaque absorption up to
$-50$ km~s$^{-1}$ and less opaque absorption beyond that to
$-250$ km~s$^{-1}$.  This difference suggests a significantly lower mass
loss rate at the time of the STIS observations.  This is consistent with
the dramatically lower UV line and continuum fluxes observed by STIS (see
Fig.~12), which suggest a lower accretion rate that would naturally be
expected to result in a lower mass loss rate (see Paper 1).

     The wind ``absorption'' features are actually a result of scattering
within the line profile.  Nevertheless, to simplify the analysis we treat
the wind features as pure absorption features.  We first approximate
the line profile above the absorption features as shown in Figure~12 for
both the STIS and IUE spectra.  The amount of flux removed by the wind
beneath these estimated background profiles is then indicative of the
amount of mass in the wind.

     We assume a radial wind with a density profile given by
\begin{equation}
\rho (r) = \frac{\dot{M}}{4\pi r^{2} V(r)},
\end{equation}
where $\dot{M}$ is the assumed mass loss rate and $V(r)$ is the wind
velocity.  Following previous analyses of wind absorption
\citep[e.g.,][]{gmh95}, we assume $V(r)$ is a power law of the form
\begin{equation}
V(r)=V_{\infty} \left( 1-\frac{R_{w}}{r} \right)^{\beta},
\end{equation}
where $V_{\infty}$ is the terminal velocity and $R_{w}$ is the radial
distance where the wind acceleration is
initiated.  In stellar wind calculations one would normally assume that
the wind is accelerated near the surface of the star and that $R_{w}$
should therefore be the radius of the star.  However, it is not known
for sure if Mira~B is a white dwarf or a red dwarf (see \S 1), so its
radius is very uncertain, but even if Mira~B's identity was clearly
known it is not certain that an accretion induced wind would originate
from close to the star.

     Our mass loss rate estimates will be proportional to the assumed
$R_{w}$ value, so the issue is not purely academic.
We will assume $R_{w}=1$ R$_{\odot}$, which seems to yield reasonable
density profiles and mass loss rates.  It seems natural to assume that the
wind is likely to originate in hot, rapidly rotating, interior regions of
the accretion disk, and we note that $R_{w}=1$ R$_{\odot}$ roughly
corresponds to the radius where \citet{dr85} place the
origin of the Si~III] $\lambda$1892 and C~III] $\lambda$1908 lines,
assuming the widths of these lines are determined by Keplerian rotation
around a 0.6~M$_{\odot}$ accretor.

     We vary the $\dot{M}$, $V_{\infty}$, and $\beta$ parameters to see
which yield the best fits to both the STIS and IUE lines in Figure~12.
In order to derive a Mg~II density profile from the total density
profile, $\rho(r)$, we assume a solar abundance for Mg \citep{ea89},
and we also assume that Mg~II is the dominant ionization state of Mg in
the wind.  Since Mira is an evolved star system, it is possible that Mg
abundances are higher than solar, in which case our mass loss rates will
be too high.  However, it is possible that a significant fraction of the
Mg could be in an ionization state other than Mg~II, in which case our
mass loss rates could also be too low.
The Mg~II flow velocity along the line of sight is
naturally provided by equation (8).  The last thing we
have to assume to compute Mg~II absorption along the line of sight is
a Doppler broadening parameter, $b$, for the absorption, which will be
characteristic of the thermal and turbulent motions of the Mg~II atoms.
We would expect thermal broadening appropriate for the temperature of
formation of Mg~II to limit $b$ to be greater than 2 km~s$^{-1}$.  When
we assume $b>6$ km~s$^{-1}$, the wind absorption broadens to velocities
greater than 0 km~s$^{-1}$, inconsistent with the absorption features
seen in Figure~12.  Thus, we believe $b=2-6$ km~s$^{-1}$, and our
results are not terribly sensitive to the assumed $b$ within this range.
Our best fits in Figure~12 assume $b=4$ km~s$^{-1}$.

     The STIS data yield a much more precise mass loss measurement than
the IUE data, because the wind absorption is not saturated.  Thus, we
focused most of our initial efforts in fitting the STIS data.  Figure~12a
shows our best fit to the STIS Mg~II profile, which assumes
$\dot{M}=5\times 10^{-13}$ M$_{\odot}$ yr$^{-1}$,
$V_{\infty}=250$ km~s$^{-1}$, and $\beta=1.15$.  Continuum fluxes and
Mg~II fluxes are about a factor of 20 higher in the IUE spectrum in
Figure~12b.  Thus, we assume $\dot{M}$ is a factor of 20 higher as well
($\dot{M}=1\times 10^{-11}$ M$_{\odot}$ yr$^{-1}$), and Figure~12b shows
that we can fit the IUE data nicely with this value if we assume
$V_{\infty}=400$ km~s$^{-1}$, with the $\beta$ parameter remaining at
$\beta=1.15$.

     Our mass loss estimates demonstrate that the wind absorption features
are entirely consistent with Mira~B's mass loss being lower by about a
factor of 20 at the time of the STIS observations relative to the IUE era,
consistent with the decrease in accretion rate suggested by the factor
of $\sim 20$ decrease in UV line and continuum flux.  Based on the data
currently available, the Mg~II profile in Figure~12b appears to be more
typical for Mira~B both in terms of its flux and its wind absorption, so
the $\dot{M}=1\times 10^{-11}$ M$_{\odot}$ yr$^{-1}$ value estimated from
this profile is probably a more typical value for Mira~B than the lower
value measured from the STIS data.  This value and the previous estimates
for Mira~B accretion rates mentioned above
($\sim 10^{-9}$ M$_{\odot}$ yr$^{-1}$) suggest that about 1\% of the
mass accreting onto Mira~B is ejected into the wind.

     Now that we have mass loss and wind velocity estimates for Mira~B, we
can estimate a location for the wind interaction region for the
Mira binary system, which will be roughly where the wind ram pressures
(i.e., $\rho V^{2}$) are equivalent.  The ratio of wind ram pressures at a
point equidistant from the two stars can be expressed as
\begin{equation}
\gamma=\frac{\dot{M}_{A} V_{A}}{\dot{M}_{B} V_{B}},
\end{equation}
where $\dot{M}_{A}$ and $V_{A}$ are the mass loss rate and terminal
velocity for Mira~A's wind, and $\dot{M}_{B}$ and $V_{B}$ are the mass loss
rate and terminal velocity for Mira~B's wind.  Equation (7) indicates how
the wind density will fall off with distance from the star, if the wind
velocities are set equal to the terminal velocities.  For $\gamma >1$, the
distance from Mira~B toward Mira~A at which the wind ram pressures will
balance is
\begin{equation}
d_{w}=\frac{1-\gamma^{0.5}}{1-\gamma} d,
\end{equation}
where $d$ is the distance between the two stars.  With a separation of
$0.578^{\prime\prime}$ and a distance from Earth of 128~pc
\citep{mk97,macp97}, the projected distance between Mira
A and B is 70~AU.  The orbit of \citet{jlp02} mentioned in \S 2.5 suggests
that the projected distance should be close to the actual distance,
so we will assume $d=70$~AU, although we note once again that the
\citet{jlp02} orbit is uncertain and anything derived from it is
therefore somewhat suspect.

     The terminal wind speed for Mira~A is roughly $V_{A}=5$ km~s$^{-1}$,
but its actual speed at the wind interaction region will be somewhat higher,
because the wind will accelerate as it falls into Mira~B's gravitational
well.  In Figure~13, we show particle trajectories for Mira~A's wind,
assuming it flows from Mira~A at a speed of $V_{A}=5$ km~s$^{-1}$ and then
is influenced only by the gravitational field of Mira~B, assuming a Mira~B
mass of 0.6~M$_{\odot}$.  The trajectories are truncated at a distance of
3.7~AU from Mira~B, because by trial and error we determine that this is
roughly the location of the wind interaction region based on the following
line of reasoning.  At this distance from Mira~B along the line toward
Mira~A, the Mira~A wind speed has increased to 13 km~s$^{-1}$.  If we
therefore assume $V_{A}=13$ km~s$^{-1}$ and
$\dot{M}_{A}=1\times 10^{-7}$ M$_{\odot}$ yr$^{-1}$,
and we assume $\dot{M}_{B}=1\times 10^{-11}$ M$_{\odot}$ yr$^{-1}$ and
$V_{B}=400$ km~s$^{-1}$ from our IUE measurements (see Fig.~12b), from
equations (9) and (10) we then compute $d_{w}=3.7$~AU, demonstrating
self-consistency with the location where we estimated $V_{A}$.

     Figure~13 indicates that all Mira~A wind material ejected within
$8^{\circ}$ of the direction of Mira~B ends up within 3.7~AU of Mira~B,
corresponding to about 0.5\% of Mira~A's total wind mass.  If this material
is accreted onto the star, the implied accretion rate is about
$5\times 10^{-10}$ M$_{\odot}$ yr$^{-1}$ (assuming
$\dot{M}_{A}=1\times 10^{-7}$ M$_{\odot}$ yr$^{-1}$).  This value is within
a factor of 2 of the $\sim 1\times 10^{-9}$ Mira~B accretion rate estimated
from its luminosity, providing support for $d_{w}=3.7$~AU being a reasonable
size scale for the wind interaction region.

     However, the accretion luminosity and Mira~B mass loss rate were
about a factor of 20 times lower at the time of our STIS observations.  If
this is due to a low density region in Mira~A's wind, then the decrease in
ram pressure of Mira~B's wind presumably balances the decrease in Mira~A's
wind, and our estimate of $d_{w}$ above does not change.  If, however, we
assume that the decrease in accretion luminosity is due to a temporary
instability rather than a change in Mira~A's wind properties,
and if we also assume that the wind interaction region has had time to
respond to the lower Mira~B mass loss rate, then our estimate of $d_{w}$
decreases to $d_{w}=0.4$~AU.  More precisely, at 0.4~AU the Mira~A wind
speed should have increased to 38 km~s$^{-1}$, so if we assume
$V_{A}=38$ km~s$^{-1}$ and
$\dot{M}_{A}=1\times 10^{-7}$ M$_{\odot}$ yr$^{-1}$,
and we assume $\dot{M}_{B}=5\times 10^{-13}$ M$_{\odot}$ yr$^{-1}$ and
$V_{B}=250$ km~s$^{-1}$ from our STIS measurements (see Fig.~12a), from
equations (9) and (10) we then compute $d_{w}=0.4$~AU.

     Thus, for the wind interaction interpretation of the H$_{2}$ emission,
we estimate that the H$_{2}$ is at a distance of roughly $0.4-3.7$~AU from
Mira~B, depending on what role variations in Mira~A's wind have played in
Mira~B's dramatic decrease in luminosity, and on whether the wind
interaction region has had time to react to the decrease in Mira~B's mass
loss rate.  However, the closer the H$_{2}$ is to Mira~B, the faster it is
expected to be moving, and therefore the closer the vector from Mira~A to
Mira~B has to be to the plane of the sky in order to explain why the
H$_{2}$ lines are not shifted significantly from the Mira rest frame
(see \S2.1).  Mostly for this reason, we suspect $d_{w}=3.7$~AU is more
likely than $d_{w}=0.4$~AU.  In any case, the size scale of the wind
interaction region is small enough that it should not be resolvable with
STIS.  The STIS aperture used for our observations has a projected size of
$24\times 24$~AU at Mira's distance, and is shown schematically in
Figure~13.

     In Figure~14, we show the spatial profile of the geocoronal H~I
Ly$\alpha$ emission (solid line), which fills the aperture and therefore has
a broad, square-topped appearance.  (The geocoronal Ly$\alpha$ line was
mentioned in Paper 1, but is removed before displaying the
Ly$\alpha$ spectrum in Figs.~7--10.)  The geocoronal profile is compared with
the spatial profile of the stellar Ly$\alpha$ emission (dotted line) and
the spatial profiles of two of the strongest H$_{2}$ features:  the 0-4 P(3)
$\lambda1342$ line and the blend of the 0-5 R(0) and 0-5 R(1) lines at
1394~\AA\ (dashed and dot-dashed lines, respectively).  There is no
significant difference between the stellar Ly$\alpha$ and H$_{2}$ profiles
in Figure~14 that would indicate that the H$_{2}$ emission is spatially
resolved, and all three profiles are significantly narrower than the
geocoronal profile.  We estimate that the radius of the H$_{2}$ emission
region can be no larger than about 6~AU to be consistent with its
unresolved nature.  Our estimates above for the wind interaction region
distance are safely below this number.

     For a Mira~A mass loss rate of $1\times 10^{-7}$ M$_{\odot}$ yr$^{-1}$
and a wind flow pattern like that in Figure~13, the Mira~A wind density
at the wind interaction region 3.7~AU from Mira~B should be about
$1.3\times 10^{-18}$ gm~cm$^{-3}$, taking into account the increase in wind
speed and the focusing effect of Mira~B's gravitational field.  If we
assume that all of this is H$_{2}$, then the implied H$_{2}$ number density
is $n(H_{2})=3.9\times 10^{5}$ cm$^{-3}$.  Our analysis in \S 2.2 suggests
a column density of $\log N(H_{2})=17.3$ for the fluoresced H$_{2}$.  The
observed H$_{2}$ emission could therefore be explained by an interaction
layer of thickness $N(H_{2})/n(H_{2})=0.03$~AU.  This value is reasonable
in that it is significantly smaller than the distance to the layer from
Mira~B, demonstrating that Mira~A's wind is dense enough at Mira~B's
location for the ``wind interaction layer'' interpretation of the H$_{2}$
emission to be valid.

     At this point, the wind interaction interpretation for the production
of the observed $T=3600$~K H$_{2}$ seems to be feasible, but problems
arise when we look at the hydrodynamics of the interaction region.
A shock must be present where the wind of Mira~A encounters the wind of
Mira~B.  A standard hydrodynamic shock could get to the correct temperature
for a shock speed of about 1 km~s$^{-1}$.  However, this speed is much lower
than one would expect given the wind speeds involved.  Furthermore, the gas
cools rapidly and one expects a column density of $\log N(H_{2})\sim 14$,
which is much too small.  Faster shocks, in better agreement with
expectations, would produce even higher temperatures and would dissociate
H$_{2}$ rapidly, leading to even lower H$_{2}$ columns.

     Hydromagnetic shocks in weakly ionized gas
\citep[i.e., C-shocks; see][]{btd93} are observed to produce temperatures
around 3000~K in Herbig-Haro objects \citep[e.g.,][]{rg96}.  It is not
difficult for such a shock to provide roughly the observed T and N(H$_{2}$),
but a strong magnetic field and at least some ionization within Mira~A's
wind are required, and it is questionable if Mira~A's cold wind can provide
these initial conditions.  Furthermore, the strong Ly$\alpha$ emission from
Mira~B will dissociate and heat the gas beyond what is found
in existing C-shock models.  This leads to a somewhat different
interpretation of the warm H$_{2}$:  a photodissociation front, in which
the H$_{2}$ still originates from within Mira~A's wind, but it is
photodissociation that heats the H$_{2}$ rather than interaction with
Mira~B's wind.  We now discuss this possibility in more detail.

\subsection{The Location of the Fluoresced H$_{2}$:  A Photodissociation
  Front?}

     The importance of photodissociation of H$_{2}$ by Ly$\alpha$ is
indicated by the last column of Table~2 ($f_{dis}$), which shows that
for some sequences as many as 40\% of fluorescence events result in
dissociation rather than the escape of an H$_{2}$ emission line photon.
Totaling up the numbers for all sequences, one finds that 20\% of all
fluorescences lead to dissociation, with the 2-1 P(13) sequence accounting
for 47\% of these dissociations.  The implied dissociation rate is
$9.7\times 10^{-3}$ cm$^{-2}$ s$^{-1}$, which at a distance of 128~pc
corresponds to $1.9\times 10^{40}$ H$_{2}$ atoms per second.  This
amounts to $1.0\times 10^{-9}$ M$_{\odot}$ yr$^{-1}$, which is roughly
Mira~B's total accretion rate (although once again, this accretion rate was
apparently lower at the time of the STIS observations).  In the previous
section we envisioned the H$_{2}$ being in a $\log N(H_{2})=17.3$ layer
about 3.7~AU from the star.  Such a layer would be completely dissociated
in less than 5 days, illustrating the need for a continual source of
H$_{2}$.  This is perhaps yet another argument in favor of Mira~A's wind
being the source of the H$_{2}$ rather than the accretion disk, since
Mira~A's massive wind apparently does typically supply about
$10^{-9}$ M$_{\odot}$ yr$^{-1}$ to Mira~B's accretion disk.

     There are two ways in which H$_{2}$ fluorescence can heat the H$_{2}$.
For the fluorescences that lead to dissociation, the fragment H~I atoms
will have substantial kinetic energy that will heat the gas.  Even the
fluorescences that do {\em not} lead to dissociation might heat
the gas since most fluorescence photons that emerge are at wavelengths
redward of the Ly$\alpha$ wavelengths where they are first fluoresced
(see, e.g., Fig.~4), meaning that the fluorescence leaves the H$_{2}$ in a
higher net excitation state than it was in before.  If this excess energy
is converted to kinetic energy, the result is a net heating of the gas.

     Time-dependent photodissociation in the ISM has been considered by
\citet{go95} and \citet{fb96}.  A crucial
difference is that H$_{2}$ in ISM photodissociation fronts can absorb
912--1100 \AA\ continuum photons in transitions from the lowest
vibrational level, so the temperature of the gas is relatively
unimportant.  The H$_{2}$ near Mira~B, however, can only absorb
Ly$\alpha$ photons in transitions that originate in levels about
1~eV above the ground, and the populations of these levels are
very sensitive to temperature.  This makes the structure highly
nonlinear.  By analogy with photoionization fronts, a 
photodissociation front could produce a weak shock on its own 
\citep[e.g.,][chapter 20]{fhs92}.

     To investigate these possibilities we have made simple numerical
models of a photodissociation front surrounding Mira~B.  We assume constant
density and speed in Mira~A's wind and simply follow the molecular fraction
and temperature of the gas under the influence of the radiation field.  We
assume that the rovibrational levels (for $v''=0-6$ and $J''=0-19$) are in
thermal equilibrium, and we compute the temperature evolution based on
radiative heating of H$_{2}$ and H$_{2}$ radiative cooling rates from
\citet{sc92}.  Forbidden C~II and O~I lines and photoelectric heating
\citep{mgw95} are also included, but are not important.

     For the H$_{2}$ heating we use the Ly$\alpha$ profile of Figure 10b
(without the absorption features) and scale to a distance of 128 pc.  For
each pumping transition in Table~2, we compute the photodissociation rate
using the $f_{dis}$ values in Table~2 and the corresponding heating based
on the fragment kinetic energies given
by \citet*{ha00}.  As mentioned above, about 80\% of
photoexcitations lead to an emergent fluorescent photon rather than
dissociation, and whenever the fluorescent photon is longward of
Ly$\alpha$, the H$_{2}$ molecule is left in a higher state than the one
which absorbed the Ly$\alpha$ photon.  Ignoring transitions to lower
vibrational levels (which are likely to be immediately reabsorbed), we use
the list of fluorescent transitions in Table~1 to estimate the average
energy difference between the initial and final states.  At low densities
that increase in energy would be simply radiated away, while at high
densities collisional deexcitation transfers the excitation energy back to
the thermal energy of the gas.  Based on the calculations of \citet{sc92}
we use a rough approximation
\begin{equation}
  q=q_{0}~\frac{n}{n+3.0\times 10^6},
\end{equation}
where $q_{0}$ is the power deposited by photoexcitation, $q$ is the
net heating of the gas, and $n$ is the H$_{2}$ number density.

     The strongest pumping transitions are somewhat optically thick, so
we compute the optical depths assuming a 10 km~s$^{-1}$ line width
and pure absorption.   We iterate the computed temperature and optical
depth structure, generally reaching convergence to a few percent in 10 
or 20 iterations.

  The temperature structure is highly nonlinear.  Once the temperature
begins to rise, the heating and dissociation rates rise rapidly until
the molecules are destroyed.  Sufficiently cold gas (less than about 400~K)
can flow to within an AU of Mira B without significant heating or
dissociation.  Gas hotter than 1000~K, on the other hand, begins to
experience a temperature rise and dissociation much farther out.  The
position of the transition depends on the balance between the flux of
molecules and the flux of dissociating photons.  It is easy to choose
parameters that give column densities much smaller than those observed, for
instance by choosing a low density or high initial temperature.
 
     Figure~15 shows a model in which the gas is relatively cool (1000~K),
so that it flows to within a few AU of Mira B before the heating becomes
rapid.  The left panel shows the kinetic temperature and molecular fraction
of hydrogen, and the right panel shows the H$_{2}$ dissociation rate and
the heating rate due to fluorescence excitation and dissociation.  Once the
temperature begins to rise, it increases rapidly, reaching 7700~K when the
gas is half dissociated.  That temperature is so high that collisional
dissociation, which is neglected in the model, becomes significant.  This
model could correspond to a flow whose temperature provides
just the right heating rate to cause a transition at 3~AU (the plot shows
part of a model that began at 1000~K at 3.05~AU), or it could
correspond to a colder flow in which a shock or other heating mechanism
suddenly raises the temperature to 1000~K at 3~AU from Mira~B.  The details
of the initial heating are unimportant provided that the gas reaches about
1000~K.  Once that happens, the excited H$_{2}$ rapidly heats and
dissociates.  Since the photodissociation region does not have a
single-valued temperature, we cannot directly compare the results of the
model with our empirically estimated values of $T=3600$ K and
$\log N(H_{2})=17.3$.  Thus, in Table~3 we compare the level populations
expressed as column densities predicted by the model with those suggested
by $T=3600$ K and $\log N(H_{2})=17.3$.  The overall agreement is quite
good, though the highest energy levels tend to be overpopulated, suggesting
that the model reaches temperatures that are too large before the molecules
are destroyed.  This could perhaps be due to the neglect of collisional
dissociation, as mentioned above.

     The model shown in Figure~15 assumes a density of
$2\times 10^{6}$ cm$^{-3}$ and a speed of 10 km~s$^{-1}$.  Based on the
flow pattern in Figure~13, the flow speed at the location of the
photodissociation front in Figure~15 should be about 14 km~s$^{-1}$,
roughly consistent with the assumed speed, and the assumed density
corresponds to a Mira~A mass loss rate of
$3\times 10^{-7}$ M$_{\odot}$ yr$^{-1}$, consistent with the range of
mass loss rates quoted for Mira~A in \S2.6.  This
combination of parameters gives column densities similar to those required
by the observations, so the strong lines are optically thick.  The
shoulders on the right sides of the photodissociation and heating rates are
typical of optically thick cases.  A photodissociation front naturally
matches the observed column densities (optical depths of a few in the strong
lines) because that condition uses all the photons available for
photodissociating the gas.

     Column densities significantly higher than those for our model are
unlikely for any photodissociation front, because once the strong lines
start to become optically thick, as they are in the model and in the
observations (see, e.g., Fig.~10), essentially all available photons will
be used for photodissociating and heating the gas.  Higher column densities
would not lead to more heating and more hot H$_{2}$.  Thus, the column
densities listed for the model in Table~3 represent a crude upper limit
for warm H$_{2}$ columns that can be produced by a photodissociation front.

     The fact that the observed warm H$_{2}$ column density is close to
this limit is a very strong point in favor of the photodissociation front
interpretation of the H$_{2}$.  In a hydrodynamic shock situation like that
discussed in \S2.6, the parameters of the shock must be tuned to obtain the
correct columns, whereas for a photodissociation front the observed
columns are a natural product of the nature of the front.  However, it is
possible to choose parameters such that column densities
are {\em lower} than this natural high density limit, where the
photodissociation is so rapid that the lines are optically thin and many
photons pass through the cool (and therefore transparent) gas outside the
front without being absorbed.

     Another very appealing aspect of the
photodissociation front interpretation is that such a front naturally
produces relative populations for the levels fluoresced by Ly$\alpha$ with
a characteristic temperature near 3500~K, because the cooler gas
contributes little to the column densities of the excited rovibrational
levels pumped by Ly$\alpha$, while the molecules have been largely
destroyed in the warmer gas.  Thus, the level populations of
the model agree fairly well with the observed populations (see Table~3).

     One possible problem with the photodissociation front interpretation
is that one must propose that Mira~A's wind has a temperature of roughly
1000~K when it enters the front.  This temperature is somewhat higher than
the temperatures one would typically expect for Mira~A's wind.  For
example, Ryde \& Sch\"{o}ier's (2001) model of Mira's circumstellar
environment suggests a temperature of about 150~K at the distance of
Mira~B.  Their model suggests, however, that strong infrared emission
from Mira~A keeps the excitation temperature of CO somewhat warmer
than this, and one might expect something similar for H$_{2}$.  Another
intriguing possibility is that the interaction between the winds of Mira~A
and Mira~B heats the H$_{2}$ enough to trigger the photodissociation
front, in which case the H$_{2}$ heating would be due to a combination of
the wind interaction envisioned in \S2.6 and the photodissociation
envisioned in this section.  Note that our model photodissociation
front is close to the $d_{w}=3.7$~AU wind interaction distance
estimated in \S2.6.  Another appealing feature of this scenario is that
the wind interaction would decelerate Mira~A's wind.  We have to assume
that the Mira~A wind flow toward Mira~B is nearly perpendicular to our
line of sight to explain why the H$_{2}$ lines that we believe are formed
in Mira~A's wind are not significantly shifted from Mira's rest
frame, but if Mira~A's wind is declerated by wind interaction before being
fluoresced and dissociated then the constraints on the orientation of the
Mira system are not as severe.  In any case, the details of the initial
heating of the H$_{2}$ are unimportant provided that the gas reaches a
temperature of about 1000~K.  Once that happens, the excited H$_{2}$
rapidly heats and dissociates.

     The code described here is meant to demonstrate the nature or the
effects of photodissociation and the associated heating.  A more detailed
model including a complete solution of the level populations, a more
detailed treatment of radiative transfer and a hydrodynamic calculation is
needed to reliably derive flow parameters.  A complete model should also
include collisional dissociation, as the temperatures exceed 3000~K.
Nevertheless, the success of the calculations in reproducing the
observed temperature and column density of the warm H$_{2}$ shows that the
most likely explanation for the observed fluorescence is that a cold wind
from Mira~A is heated within a distance of a few AU from Mira~B by a
photodissociation front.  Any initial heating is rapidly followed by
photoexcitation and photodissociation heating, so the layer of warm H$_{2}$
is very thin.

     As an aside, a nonlinear photodissociation front similar to those
described here might operate in other situations where a strong, broad
Ly$\alpha$ emission line dominates the UV radiation field.  The faint
H$_{2}$ infrared emission features well ahead of the main shocks in such
Herbig-Haro objects as Cepheus A \citep{ph00} or HH2
\citep{jcr97}, and supernova remnants such as the Cygnus
Loop \citep{jrg91} or RCW 103 \citep{eo99}, might be
signatures of the heating in a time-dependent photodissociation front.
However, especially in the supernova remnant cases, photoionization may play
an important role, leading to structures more closely related to those
discussed by \citet{fb96}.

\subsection{Mira~B's Variability Revisited}

     Perhaps the biggest question about the STIS observations of Mira~B
concerns the fundamental cause of the order of magnitude drop in accretion
luminosity compared to the IUE era (see \S 1).  One possibility is that
Mira~B has entered a region of low density within Mira~A's wind, resulting
in Mira~A's wind supplying an order of magnitude less material to Mira~B's
accretion disk.  There is evidence for substantial inhomogeneity in
Mira~A's massive wind \citep{bl97,mm01}.  However,
this interpretation is difficult to reconcile with the fact that Mira~B's
Ly$\alpha$ emission appears to be dissociating about
$1.0\times 10^{-9}$ M$_{\odot}$ yr$^{-1}$ worth of H$_{2}$.  This
corresponds with Mira~B's normal accretion rate, which in \S2.6 we showed
was roughly consistent with the amount of material that approaches within
about 4~AU of Mira~B.  However, if Mira~A's wind density is assumed to be
lower by an order of magnitude, the interaction region within the wind
required to supply $1.0\times 10^{-9}$ M$_{\odot}$ yr$^{-1}$ of H$_{2}$ is
much larger, which is inconsistent with the H$_{2}$ emission being
unresolved and within $\sim 6$~AU of Mira~B (see \S2.6).  The recent
infrared observations of \citet{mm01} actually suggest an
{\em excess} of circumstellar material in the vicinity of Mira~B rather
than a deficit, although perhaps this could actually be a semi-permanent
excess due to the gravitational focusing effect suggested by Figure~13.

     A second possible reason for Mira~B's dramatic drop in luminosity is
that Mira~A's wind is still supplying the usual amount of H$_{2}$ to the
region of Mira~B, but Mira~B is simply not accreting as much of it as
usual, perhaps due to an accretion instability.  More luminous
accretion systems, such as dwarf novae and cataclysmic variables, show
dramatic variability due to such instabilities.  This explanation currently
seems the more likely possibility since it does not have the dissociation
rate problem that the variable wind interpretation has.

     Another issue concerns why the H$_{2}$ emission suddenly came to
dominate Mira~B's UV spectrum when the accretion luminosity dropped.
The H$_{2}$ flux fluoresced by Ly$\alpha$ is naturally related to the
amount of Ly$\alpha$ flux, so in Figure~16 we compare the STIS Ly$\alpha$
profile with a coaddition of three IUE observations of the Ly$\alpha$ line
(SWP9953, SWP20420, and SWP40597).  These are the only three high
resolution spectra containing Ly$\alpha$ taken by IUE.  The spectra are
extracted from the archive, having been processed using the standard
NEWSIPS software \citep{jsn96}.  Low resolution IUE spectra of
Ly$\alpha$ are useless since the stellar Ly$\alpha$ flux is completely
blended with the geocoronal Ly$\alpha$ line.

     The strong geocoronal Ly$\alpha$ line is also a serious problem in the
high resolution IUE spectra.  Scattered geocoronal light makes
background subtraction in that spectral region very difficult.  Inaccurate
background subtraction is undoubtedly responsible for the negative fluxes
seen below 1216.5~\AA\ in Figure~16.  These problems make it very
difficult to study weak Ly$\alpha$ emission with IUE, and we cannot even
be truly confident that the generally positive flux above 1216.5~\AA\
accurately represents the actual stellar Ly$\alpha$ flux level.  The
situation is much better for STIS.  The geocoronal emission is much
narrower and weaker thanks to the smaller aperture used by HST/STIS, and
the signal-to-noise level of the data is much higher (see Fig.~16).

     Despite the difficulties in trusting IUE Ly$\alpha$ spectra, there
is one important statement that we believe we can confidently make based
on Figure~16:  Unlike the continuum and all other non-H$_{2}$ lines,
the Ly$\alpha$ line observed by STIS is {\em not} over an order of
magnitude weaker than it was during the IUE era.  Figure~16 suggests that
the Ly$\alpha$ flux from Mira~B in the IUE era was at most roughly equal
to the flux we observe with STIS.  This is perhaps the primary reason
why the character of the STIS FUV spectrum is so different from that of
the IUE spectra.  Unlike the other emission lines, the Ly$\alpha$ line is
at least as strong as during the IUE era, if not stronger, so the H$_{2}$
emission is correspondingly much stronger relative to the other emission
lines.

     This leads naturally to the question of why the Ly$\alpha$ behavior
is so different from that of the other emission lines.  It is difficult to
imagine why an order of magnitude decrease in Mg~II flux from Mira~B (see
Fig.~12) would not also be accompanied by an order of magnitude decrease in
Ly$\alpha$ flux.  The Mg~II and Ly$\alpha$ lines have similar line
formation temperatures and both are affected by wind absorption.  However,
the opacity of the wind to Ly$\alpha$ emission will be many orders of
magnitude higher than for Mg~II, so perhaps the solution lies in this
fundamental difference.

     Exactly how this might work is not clear, though, since
wind absorption features like those in Figure~12 are generally assumed
to be a result of scattering rather than pure absorption.  Scattering
merely changes the wavelengths of photons within the line profile rather
than destroying the photons, but we require a mechanism that would destroy
Ly$\alpha$ photons but not Mg~II photons.  Thus, perhaps there is a photon
destruction mechanism that is operating on the Ly$\alpha$ photons, although
we can only speculate as to what it might be.  Perhaps the higher opacities
of the wind during the IUE era scatter the Ly$\alpha$ photons back into the
dense accretion disk where they are thermalized, while
Mg~II opacities are still low enough to allow Mg~II photons to ultimately
escape.

\section{SUMMARY}

     We have measured and analyzed 103 H$_{2}$ lines in a far-UV HST/STIS
spectrum of Mira~B.  These lines are all fluoresced by H~I Ly$\alpha$
emission, and we have divided the lines into 13 different fluorescence
sequences, representing 13 H$_{2}$ transitions within the Ly$\alpha$ line
that are being pumped by the Ly$\alpha$ emission.  Detailed analysis of
the H$_{2}$ emission has led to the following conclusions:
\begin{description}
\item[1.] The average centroid velocity of the H$_{2}$ lines is
  $56.9\pm 0.2$ km~s$^{-1}$, consistent with the $\sim 56$ km~s$^{-1}$
  radial velocity of the Mira binary.  The average width of the H$_{2}$
  lines is $FWHM=19.7\pm 0.4$ km~s$^{-1}$.
\item[2.] The observed flux ratios within the H$_{2}$ fluorescence
  sequences show evidence for opacity effects that cause the low wavelength
  lines to systematically have lower fluxes than the line branching ratios
  predict, due to the higher opacity of those lines.  We use a Monte Carlo
  radiative transfer code to model the H$_{2}$ line ratios that emerge
  from a plane parallel slab.  We find that the observed line ratios are
  best reproduced by a slab with temperature $T=3600$~K and column density
  $\log {\rm N(H_{2})}=17.3$.
\item[3.] Even for our best fit to the data, there are still significant
  discrepancies between the observed H$_{2}$ line ratios and those
  predicted by our radiative transfer code.  The discrepancies are largest
  for lines in sequences fluoresced from lower energy levels, suggesting
  that the populations of these lower levels may not be entirely determined
  by thermal processes.
\item[4.] We find that we can use the total flux fluoresced by each of the
  13 H$_{2}$ pumping transitions within Ly$\alpha$ to map out a
  self-consistent Ly$\alpha$ profile seen by the H$_{2}$, assuming the
  temperature of $T=3600$~K derived from the line ratio analysis.  This
  means that the H$_{2}$ energy levels can at least to first order be
  described by a thermal population appropriate for $T=3600$~K.  When we
  change the assumed temperature, the analysis is not as successful
  in deriving a smooth, self-consistent Ly$\alpha$ profile, providing
  support for the $T=3600$~K measurement.
\item[5.] The observed Ly$\alpha$ profile is completely redshifted from the
  rest frame of the star, presumably due to wind absorption eating away the
  blue side of the line.  The model Ly$\alpha$ profile seen by the H$_{2}$
  that we derive is also completely redshifted relative to the stellar
  rest frame, implying that the wind absorption must occur before the
  Ly$\alpha$ emission encounters the H$_{2}$.
\item[6.] The fluxes of the model Ly$\alpha$ profile seen by the H$_{2}$
  are significantly higher than those of the observed Ly$\alpha$ profile.
  Interstellar and/or circumstellar H~I absorption of Ly$\alpha$ can
  account for much, but not all, of this discrepancy.  We find that we
  can best reconcile the modeled and observed Ly$\alpha$ profiles if we
  assume an interstellar (or circumstellar) H~I column density of
  $\log {\rm N(H~I)}=20.35$, and if we also assume that Ly$\alpha$ fluxes
  along our line of sight are suppressed by a factor of 2.5 compared to
  the average fluxes seen by the H$_{2}$.  We speculate that perhaps this
  suppression factor of 2.5 is due to a near edge-on orientation of
  Mira~B's accretion disk relative to our line of sight --- either dust
  extinction from the disk or higher H~I wind densities near the disk could
  act to suppress the observed Ly$\alpha$ fluxes for our line of sight.
\item[7.] We observe H$_{2}$ absorption features within the Ly$\alpha$ line
  at the location of the transitions where the H$_{2}$ is being fluoresced.
  The strength of the absorption is roughly consistent with the amount of
  absorption predicted by an H$_{2}$ slab with $T=3600$~K and
  $\log {\rm N(H_{2})}=17.3$, providing further support for the accuracy of
  these measurements.
\item[8.] We propose that the most likely location for the H$_{2}$ that is
  being fluoresced is in the wind of Mira~A. The H$_{2}$ absorption visible
  within the Ly$\alpha$ line, our finding that the H$_{2}$ most likely
  completely surrounds the star, and the lack of rotational broadening
  signatures in the H$_{2}$ line profiles argue against the H$_{2}$ being
  within the accretion disk.
\item[9.] One possibility is that the H$_{2}$ that we observe is within
  the interaction region of the winds of Mira~A and Mira~B.  Thus, we
  estimate the properties of Mira~B's wind from wind absorption observed
  within the Mg~II h \& k lines.  Typical IUE Mg~II profiles suggest
  mass loss rates of $\sim 1\times 10^{-11}$ M$_{\odot}$ yr$^{-1}$ and
  terminal speeds of $V_{\infty}=400$ km~s$^{-1}$, while for our more
  recent STIS data we find $\dot{M}=5\times 10^{-13}$ M$_{\odot}$ yr$^{-1}$
  and a terminal velocity $V_{\infty}=250$ km~s$^{-1}$.  The lower STIS
  mass loss rate is consistent with the dramatically lower UV line and
  continuum fluxes seen by STIS compared with IUE, corresponding to a
  lower accretion rate and naturally a lower mass loss rate.
\item[10.] Based on our estimates for Mira~B's mass loss rate from the
  Mg~II analysis, and based on previously published estimates for Mira~A's
  wind properties, we estimate that the ram pressures of the two winds
  will balance about 3.7~AU from Mira~B.  This is one possible location for
  the warm H$_{2}$ that we observe, but simple hydrodynamic shock
  calculations have difficulty reproducing the observed H$_{2}$ temperature
  and column density.
\item[11.] We use crude models to demonstrate that a photodissociation
  front has more success reproducing the properties of the H$_{2}$ emission
  that we observe than a wind interaction shock, since photodissociation
  fronts naturally tend to yield temperatures of $\sim 3600$~K and
  warm H$_{2}$ columns of $\log N(H_{2})\sim 17.3$, consistent with
  observations.  It is possible, however, that the H$_{2}$ could be
  produced in a combination of a wind interaction shock and a
  photodissociation front.
\item[12.] One possible explanation for the order of magnitude decrease in
  Mira~B's accretion luminosity from the IUE era to that of our STIS
  observations is a decrease in Mira~A's wind density near Mira~B.
  However, the H$_{2}$ lines suggest a large H$_{2}$ dissociation rate of
  $1.0\times 10^{-9}$ M$_{\odot}$ yr$^{-1}$, emphasizing the need for
  a constant supply of H$_{2}$.  Since it is doubtful that a weaker
  Mira~A wind could supply this amount of H$_{2}$, we propose that
  accretion instabilities are a more likely cause of the
  drop in accretion luminosity.
\item[13.] The Ly$\alpha$ flux observed from Mira~B in our STIS
  data is {\em not} lower than the Ly$\alpha$ flux observed in the IUE era,
  in contrast with the UV continuum and all other non-H$_{2}$ lines, which
  are all lower by over an order of magnitude.  We believe this is the
  primary reason why the H$_{2}$ lines fluoresced by Ly$\alpha$ are far
  more prominent in the STIS data, while being undetectable in the IUE
  spectra.  It is uncertain why the Ly$\alpha$ fluxes did not change with
  the other UV line and continuum fluxes, but we speculate that perhaps
  the stronger wind present in the IUE era somehow suppressed the
  Ly$\alpha$ flux.
\end{description}

\acknowledgments

Support for this work was provided by NASA through grant number
GO-08298.01-99A from the Space Telescope Science Institute, which is operated
by AURA, Inc., under NASA contract NAS5-26555.  M.\ K.\ is a member of the
Chandra Science Center, which is operated under contract NAS8-39073,
and is partially supported by NASA.


\topmargin -0.2in
\textheight 9.2in

\begin{deluxetable}{lcccclccc}
\tabletypesize{\tiny}
\tablecaption{Detected H$_{2}$ Lines}
\tablecolumns{9}
\tablewidth{0pt}
\tablehead{
  \colhead{ID} & \colhead{$\lambda_{rest}$} & 
    \multicolumn{2}{c}{Fluxes ($10^{-15}$)} & &
  \colhead{ID} & \colhead{$\lambda_{rest}$} & 
    \multicolumn{2}{c}{Fluxes ($10^{-15}$)} \\
  \colhead{} & \colhead{(\AA)} & \colhead{Observed} & \colhead{Model} & &
  \colhead{} & \colhead{(\AA)} & \colhead{Observed} & \colhead{Model}}
\startdata
\multicolumn{4}{c}{\underline{Fluoresced by 1-2 R(6) $\lambda$1215.7263}} & &
 \multicolumn{4}{c}{\underline{Fluoresced by 2-1 P(13) $\lambda$1217.9041}} \\
1-2 P(8)   &    1237.8623   &    $1.70\pm  0.39$   &    1.16 & &
  2-2 P(13)  &     1271.1773  &    $4.18\pm  0.60$   &    3.36 \\
1-3 P(8)   &    1293.8674   &    $2.85\pm  0.42$   &    2.63 & &
  2-3 P(13)  &     1325.3415  &    $1.21\pm  0.30$   &    1.85 \\
1-6 P(8)   &    1467.0791   &    $4.15\pm  0.56$   &    4.55 & &
  2-4 P(13)  &     1379.9818  &    $1.91\pm  0.41$   &    2.09 \\
1-7 P(8)   &    1524.6484   &    $5.99\pm  0.74$   &    6.45 & &
  2-5 P(13)  &     1434.5327  &    $6.44\pm  0.63$   &    8.35 \\
1-8 P(8)   &    1580.6660   &    $4.24\pm  0.88$   &    3.73 & &
  2-6 P(13)  &     1488.2389  &    $4.11\pm  0.68$   &    2.80 \\
1-3 R(6)   &    1271.0135   &    $2.42\pm  0.47$   &    2.07 & &
  2-7 P(13)  &     1540.0937  &    $3.21\pm  0.59$   &    2.52 \\
1-4 R(6)   &    1327.5595   &    $0.58\pm  0.35$   &    1.43 & &
  2-8 P(13)  &     1588.7925  &    $15.49\pm 1.39$   &    14.91 \\
1-6 R(6)   &    1442.8602   &    $2.87\pm  0.45$   &    3.21 & &
  2-9 P(13)  &     1632.6080  &    $8.90\pm  1.36$   &    9.24 \\
1-7 R(6)   &    1500.4424   &    $6.97\pm  0.81$   &    6.11 & &
  2-2 R(11)  &     1237.5357  &    $3.13\pm  0.50$   &    1.81 \\
1-8 R(6)   &    1556.8600   &    $4.97\pm  0.71$   &    4.28 & &
  2-3 R(11)  &     1290.8972  &    $3.47\pm  0.41$   &    3.33 \\
 & & & & &
  2-5 R(11)  &     1399.2344  &    $6.50\pm  0.64$   &    6.36 \\
\multicolumn{4}{c}{\underline{Fluoresced by 1-2 P(5) $\lambda$1216.0696}} & &
  2-6 R(11)  &     1453.0927  &    $6.34\pm  0.63$   &    6.26 \\
1-3 P(5)   &    1271.9246   &    $2.99\pm  0.46$   &    1.90 & &
  2-8 R(11)  &     1555.8804  &    $9.85\pm  0.98$   &    9.92 \\
1-5 P(5)   &    1387.3618   &    $1.57\pm  0.36$   &    1.57 & &
  2-9 R(11)  &     1602.2641  &    $16.48\pm 1.70$   &    13.93 \\
1-6 P(5)   &    1446.1175   &    $9.10\pm  0.86$   &    9.10 & &
  2-10 R(11) &     1642.9363  &    $4.76\pm  1.04$   &    2.72 \\
1-7 P(5)   &    1504.7509   &    $15.64\pm 1.17$   &    14.65 & &
  & & & \\
1-8 P(5)   &    1562.3886   &    $8.74\pm  0.98$   &    8.76 & &
 \multicolumn{4}{c}{\underline{Fluoresced by 2-1 R(14) $\lambda$1218.5205}} \\
1-3 R(3)   &    1257.8277   &    $2.47\pm  0.47$   &    1.43 & &
  2-1 P(16)  &     1257.3939  &    $1.27\pm  0.29$   &    1.39 \\
1-4 R(3)   &    1314.6130   &    $1.76\pm  0.37$   &    1.72 & &
  2-2 P(16)  &     1310.5466  &    $2.14\pm  0.39$   &    2.19 \\
1-6 R(3)   &    1431.0102   &    $6.34\pm  0.76$   &    6.53 & &
  2-5 P(16)  &     1470.4409  &    $2.40\pm  0.54$   &    2.03 \\
1-7 R(3)   &    1489.5637   &    $10.70\pm 0.97$   &    12.30 & &
  2-8 P(16)  &     1612.3812  &    $4.63\pm  1.03$   &    4.00 \\
1-8 R(3)   &    1547.3337   &    $6.93\pm  0.92$   &    7.98 & &
  2-2 R(14)  &     1270.7436  &    $1.77\pm  0.75$   &    2.21 \\
1-9 R(3)   &    1603.2490   &    $4.73\pm  1.01$   &    2.65 & &
  2-3 R(14)  &     1323.6694  &    $1.09\pm  0.31$   &    1.49 \\
 & & & & &
  2-5 R(14)  &     1429.7062  &    $1.85\pm  0.42$   &    1.64 \\
\multicolumn{4}{c}{\underline{Fluoresced by 3-3 P(1) $\lambda$1217.0377}} & &
  2-6 R(14)  &     1481.4176  &    $1.27\pm  0.32$   &    1.54 \\
3-7 P(1)   &    1435.0477   &    $1.05\pm  0.32$   &    1.56 & &
  2-8 R(14)  &     1576.8729  &    $4.39\pm  0.74$   &    2.92 \\
3-10 P(1)  &    1591.3101   &    $4.72\pm  0.83$   &    2.82 & &
  & & & \\
 & & & & &
 \multicolumn{4}{c}{\underline{Fluoresced by 0-2 R(2) $\lambda$1219.0887}} \\
\multicolumn{4}{c}{\underline{Fluoresced by 0-2 R(0) $\lambda$1217.2045}} & &
  0-3 P(4)   &    1287.7306   &    $1.93\pm  0.34$   &    1.02 \\
0-3 P(2)   &    1279.4639   &    $4.08\pm  0.45$   &    7.49 & &
  0-4 P(4)   &    1346.9081   &    $2.27\pm  0.35$   &    2.08 \\
0-4 P(2)   &    1338.5682   &    $16.33\pm 0.81$   &    13.08 & &
  0-5 P(4)   &    1407.2856   &    $3.93\pm  0.50$   &    2.31 \\
0-5 P(2)   &    1398.9514   &    $17.08\pm 1.02$   &    12.77 & &
  0-6 P(4)   &    1468.3873   &    $2.37\pm  0.46$   &    1.53 \\
0-6 P(2)   &    1460.1650   &    $11.48\pm 0.97$   &    8.23 & &
  0-3 R(2)   &    1276.3223   &    $0.71\pm  0.26$   &    0.79 \\
0-7 P(2)   &    1521.5869   &    $4.07\pm  0.72$   &    3.26 & &
  0-4 R(2)   &    1335.1300   &    $0.88\pm  0.27$   &    1.62 \\
0-3 R(0)   &    1274.5343   &    $4.42\pm  0.51$   &    4.74 & &
  0-5 R(2)   &    1395.1965   &    $1.20\pm  0.29$   &    1.88 \\
0-4 R(0)   &    1333.4741   &    $6.44\pm  0.57$   &    7.42 & &
  & & & \\
0-5 R(0)   &    1393.7190   &    $8.11\pm  0.73$   &    7.20 & &
  \multicolumn{4}{c}{\underline{Fluoresced by 2-2 R(9) $\lambda$1219.1005}} \\
0-6 R(0)   &    1454.8287   &    $3.46\pm  0.59$   &    4.24 & &
  2-2 P(11)  &    1248.1447   &    $1.24\pm  0.33$   &    0.52 \\
0-7 R(0)   &    1516.1972   &    $2.28\pm  0.68$   &    1.79 & &
  2-5 P(11)  &    1412.8123   &    $1.32\pm  0.37$   &    2.23 \\
 & & & & &
  2-8 P(11)  &    1572.6868   &    $4.73\pm  0.79$   &    4.05 \\
\multicolumn{4}{c}{\underline{Fluoresced by 4-0 P(19) $\lambda$1217.4100}}& &
  2-9 P(11)  &    1620.0932   &    $3.77\pm  1.05$   &    3.18 \\
4-1 P(19)  &     1266.8791  &    $2.23\pm  0.39$   &    2.04 & &
  2-5 R(9)   &    1381.9545   &    $1.63\pm  0.35$   &    1.81 \\
4-4 P(19)  &     1414.6960  &    $1.52\pm  0.33$   &    1.30 & &
  2-6 R(9)   &    1436.8059   &    $1.95\pm  0.39$   &    1.64 \\
4-2 R(17)  &     1274.0330  &    $1.46\pm  0.34$   &    1.45 & &
  2-9 R(9)   &    1592.1893   &    $4.84\pm  0.84$   &    3.91 \\
4-4 R(17)  &     1372.0611  &    $1.20\pm  0.35$   &    1.90 & &
  & & & \\
4-6 R(17)  &     1465.1862  &    $1.98\pm  0.41$   &    1.30 & &
  \multicolumn{4}{c}{\underline{Fluoresced by 2-2 P(8) $\lambda$1219.1543}} \\
 & & & & &
  2-5 P(8)   &    1384.7720   &    $1.80\pm  0.34$   &    2.60 \\
\multicolumn{4}{c}{\underline{Fluoresced by 0-2 R(1) $\lambda$1217.6426}} & &
  2-6 P(8)   &    1440.8746   &    $1.48\pm  0.42$   &    0.97 \\
0-3 P(3)   &    1283.1111   &    $5.89\pm  0.62$   &    4.87 & &
  2-8 P(8)   &    1550.2888   &    $3.78\pm  0.90$   &    3.75 \\
0-4 P(3)   &    1342.2559   &    $18.88\pm 1.00$   &    13.55 & &
  2-9 P(8)   &    1601.3989   &    $5.52\pm  1.27$   &    3.68 \\
0-5 P(3)   &    1402.6484   &    $25.39\pm 1.11$   &    21.30 & &
  2-5 R(6)   &    1361.6294   &    $2.51\pm  0.42$   &    2.22 \\
0-6 P(3)   &    1463.8260   &    $18.59\pm 1.06$   &    16.34 & &
  2-6 R(6)   &    1417.5063   &    $1.90\pm  0.42$   &    1.55 \\
0-7 P(3)   &    1525.1532   &    $7.50\pm  0.76$   &    7.19 & &
  2-8 R(6)   &    1527.3823   &    $1.90\pm  0.48$   &    2.57 \\
0-3 R(1)   &    1274.9215   &    $5.96\pm  0.58$   &    4.24 & &
  2-9 R(6)   &    1579.4050   &    $5.40\pm  0.87$   &    3.79 \\
0-4 R(1)   &    1333.7974   &    $6.08\pm  0.55$   &    10.74 & &
  2-10 R(6)  &    1627.6722   &    $2.48\pm  0.67$   &    1.39 \\
0-5 R(1)   &    1393.9613   &    $14.70\pm 0.92$   &    15.98 & &
  & & & \\
0-6 R(1)   &    1454.9710   &    $10.21\pm 0.92$   &    11.92 & &
  \multicolumn{4}{c}{\underline{Fluoresced by 0-2 P(1) $\lambda$1219.3676}} \\
0-7 R(1)   &    1516.2181   &    $5.27\pm  1.58$   &    5.46 & &
  0-3 P(1)   &    1276.8127   &    $4.01\pm  0.49$   &    3.46 \\
 & & & & &
  0-4 P(1)   &    1335.8675   &    $8.94\pm  0.71$   &    7.12 \\
 & & & & &
  0-5 P(1)   &    1396.2225   &    $5.67\pm  0.55$   &    7.97 \\
 & & & & &
  0-6 P(1)   &    1457.4346   &    $6.97\pm  0.63$   &    5.38 \\
 & & & & &
  0-7 P(1)   &    1518.8937   &    $3.63\pm  0.78$   &    2.07 \\
 & & & & &
  & & & \\
 & & & & &
  \multicolumn{4}{c}{\underline{Fluoresced by 0-1 R(11) $\lambda$1219.7454}}\\
 & & & & &
  0-5 P(13)  &     1485.4157  &    $1.86\pm  0.46$   &    2.04 \\
 & & & & &
  0-3 R(11)  &     1331.9550  &    $0.65\pm  0.26$   &    0.70 \\
 & & & & &
  0-4 R(11)  &     1389.5844  &    $1.79\pm  0.46$   &    1.48 \\
\enddata
\end{deluxetable}

\clearpage

\begin{deluxetable}{cccccccc}
\tabletypesize{\small}
\tablecaption{Fluorescence Sequences}
\tablecolumns{8}
\tablewidth{0pt}
\tablehead{
  \colhead{Fluorescing} & \colhead{$\lambda_{rest}$} & \colhead{$F_{obs}$} &
    \colhead{Absorption} & \colhead{$E_{low}$} & \colhead{$g_{low}$} &
    \colhead{$f_{dis}^{\prime}$} & \colhead{$f_{dis}$} \\
  \colhead{Transition} & \colhead{(\AA)} & \colhead{($10^{-13}$)} & 
    \colhead{Strength (f)} & \colhead{(cm$^{-1}$)} & \colhead{} & \colhead{}}
\startdata
1-2 R(6)  & 1215.7263 & 0.467 & 0.0349 & 10261.20 &  13 & 0.170 & 0.218 \\
1-2 P(5)  & 1216.0696 & 0.865 & 0.0289 &  9654.15 &  33 & 0.001 & 0.002 \\
3-3 P(1)  & 1217.0377 & 0.111 & 0.0013 & 11883.51 &   9 & 0.0 & 0.0 \\
0-2 R(0)  & 1217.2045 & 0.831 & 0.0441 &  8086.93 &   1 & 0.0 & 0.0 \\
4-0 P(19) & 1217.4100 & 0.178 & 0.0093 & 17750.25 & 117 & 0.379 & 0.427 \\
0-2 R(1)  & 1217.6426 & 1.295 & 0.0289 &  8193.81 &   9 & 0.0 & 0.0 \\
2-1 P(13) & 1217.9041 & 1.084 & 0.0192 & 13191.06 &  81 & 0.337 & 0.407 \\
2-1 R(14) & 1218.5205 & 0.316 & 0.0181 & 14399.08 &  29 & 0.358 & 0.384 \\
0-2 R(2)  & 1219.0887 & 0.159 & 0.0256 &  8406.29 &   5 & 0.0 & 0.0 \\
2-2 R(9)  & 1219.1005 & 0.300 & 0.0318 & 12584.80 &  57 & 0.290 & 0.362 \\
2-2 P(8)  & 1219.1543 & 0.326 & 0.0214 & 11732.12 &  17 & 0.167 & 0.212 \\
0-2 P(1)  & 1219.3676 & 0.303 & 0.0149 &  8193.81 &   9 & 0.0 & 0.0 \\
0-1 R(11) & 1219.7454 & 0.163 & 0.0037 & 10927.12 &  69 & 0.344 & 0.420 \\
\enddata
\end{deluxetable}

\begin{deluxetable}{ccccc}
\tabletypesize{\small}
\tablecaption{Level Population Comparison}
\tablecolumns{5}
\tablewidth{0pt}
\tablehead{
  \colhead{$v''$} & \colhead{$J''$} &
    \colhead{$[\log N_{pop}]_{obs}$\tablenotemark{a}} &
    \colhead{$[\log N_{pop}]_{mod}$\tablenotemark{b}} &
    \colhead{Model/Obs.}}
\startdata
2 & 6 & 14.572 & 14.501 & 0.85 \\
2 & 5 & 15.082 & 15.013 & 0.85 \\
3 & 1 & 14.131 & 14.127 & 1.00 \\
2 & 0 & 13.836 & 13.894 & 1.14 \\
0 &19 & 14.226 & 14.732 & 3.21 \\
2 & 1 & 14.771 & 14.814 & 1.11 \\
1 &13 & 14.858 & 14.946 & 1.22 \\
1 &14 & 14.202 & 14.391 & 1.54 \\
2 & 2 & 14.479 & 14.496 & 1.04 \\
2 & 9 & 14.811 & 14.853 & 1.10 \\
2 & 8 & 14.433 & 14.420 & 0.97 \\
2 & 1 & 14.771 & 14.814 & 1.11 \\
1 &11 & 15.182 & 15.127 & 0.88 \\
\enddata
\tablenotetext{a}{Level populations (expressed as column densities) based
  on the empirical measurements of $T=3600$~K and $\log N(H_{2})=17.3$.}
\tablenotetext{b}{Level populations (expressed as column densities) based
  on a model photodissociation front.}
\end{deluxetable}

\clearpage

\begin{figure}
\plotfiddle{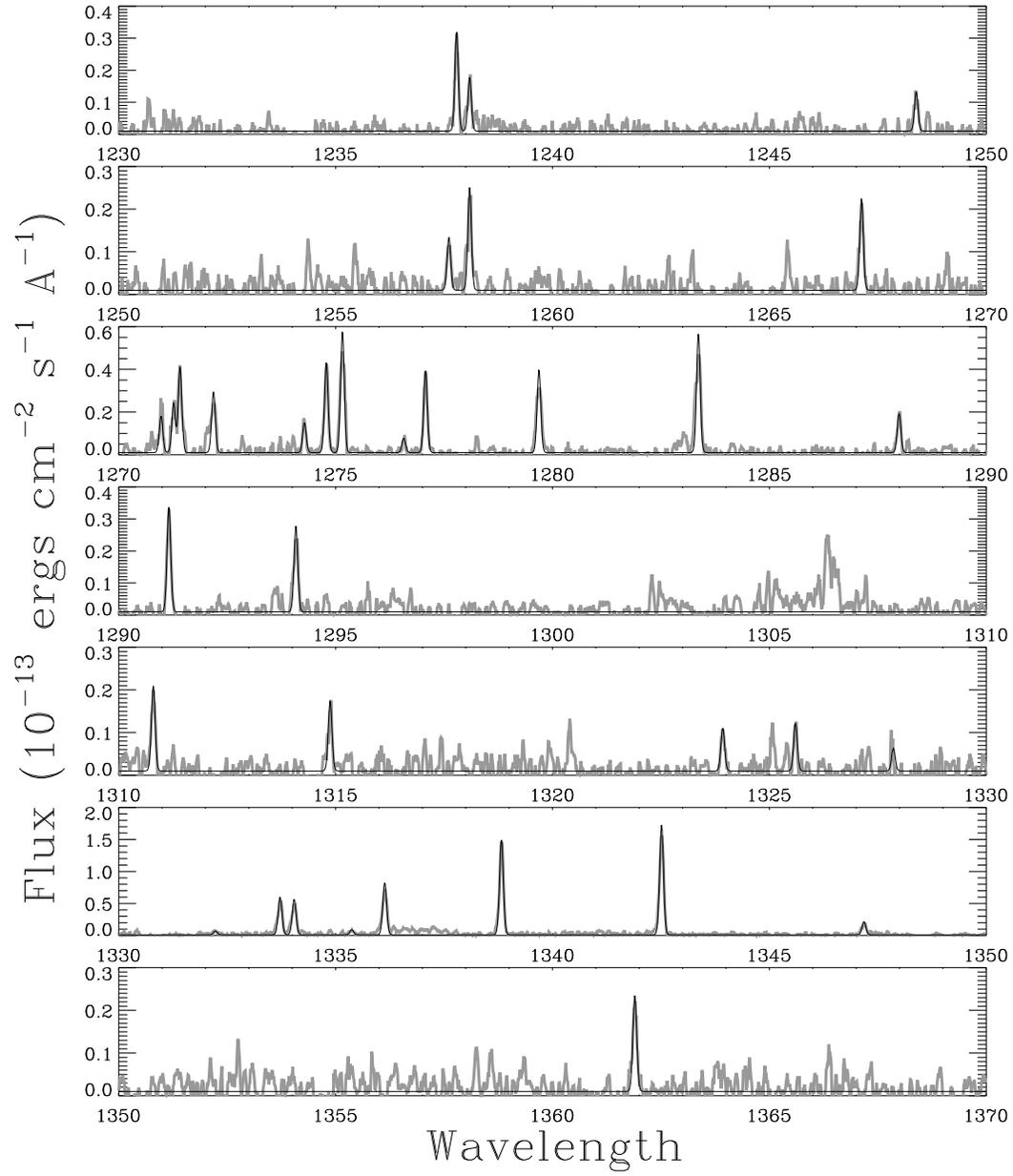}{6.5in}{0}{80}{80}{-340}{10}
\caption{The E140M STIS spectrum of Mira~B, showing
  numerous narrow H$_{2}$ lines, which are all fitted with Gaussians (thin
  solid lines).}
\end{figure}

\clearpage

\setcounter{figure}{0}
\begin{figure}
\plotfiddle{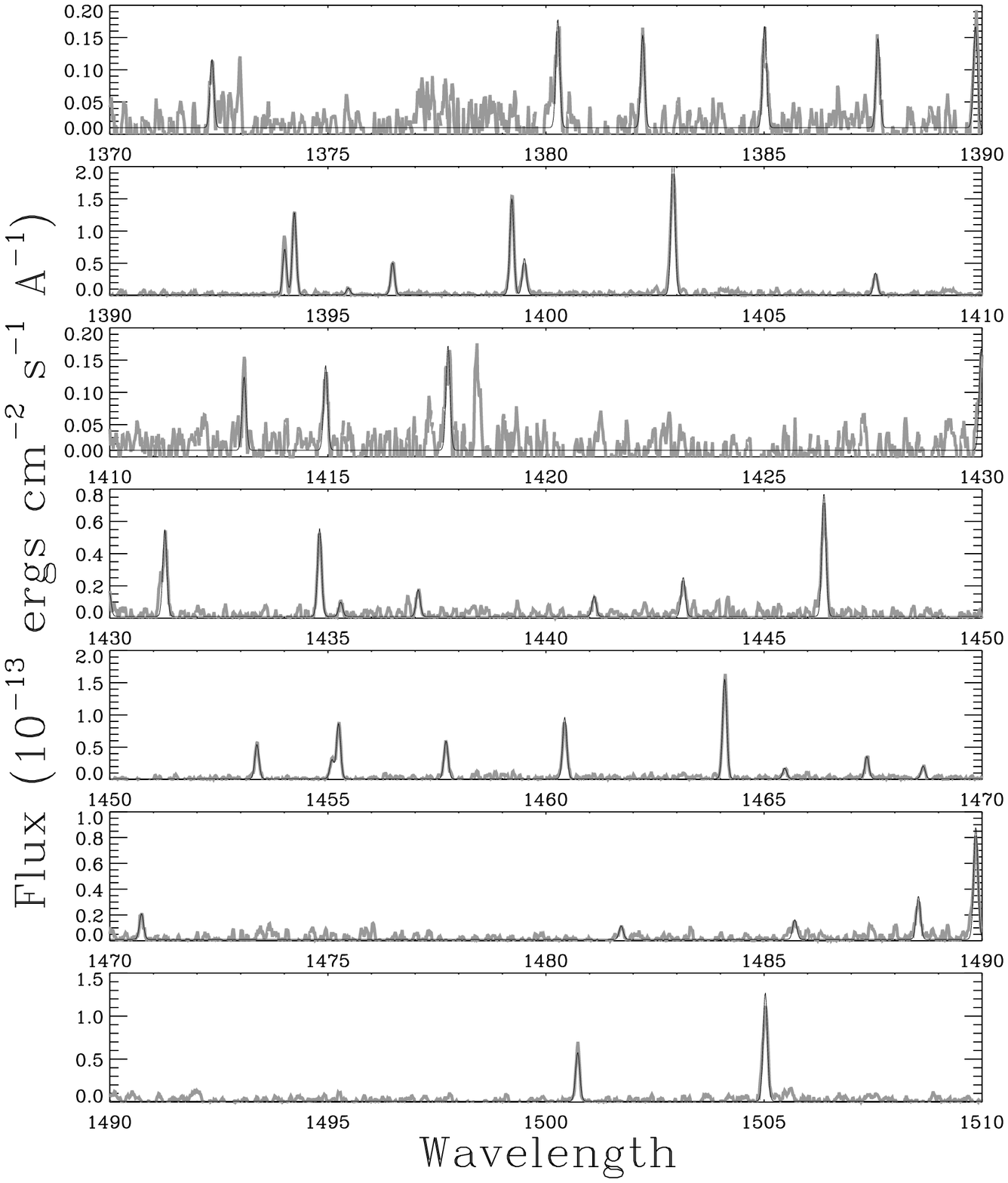}{6.5in}{0}{80}{80}{-340}{10}
\caption{(continued)}
\end{figure}

\clearpage

\setcounter{figure}{0}
\begin{figure}
\plotfiddle{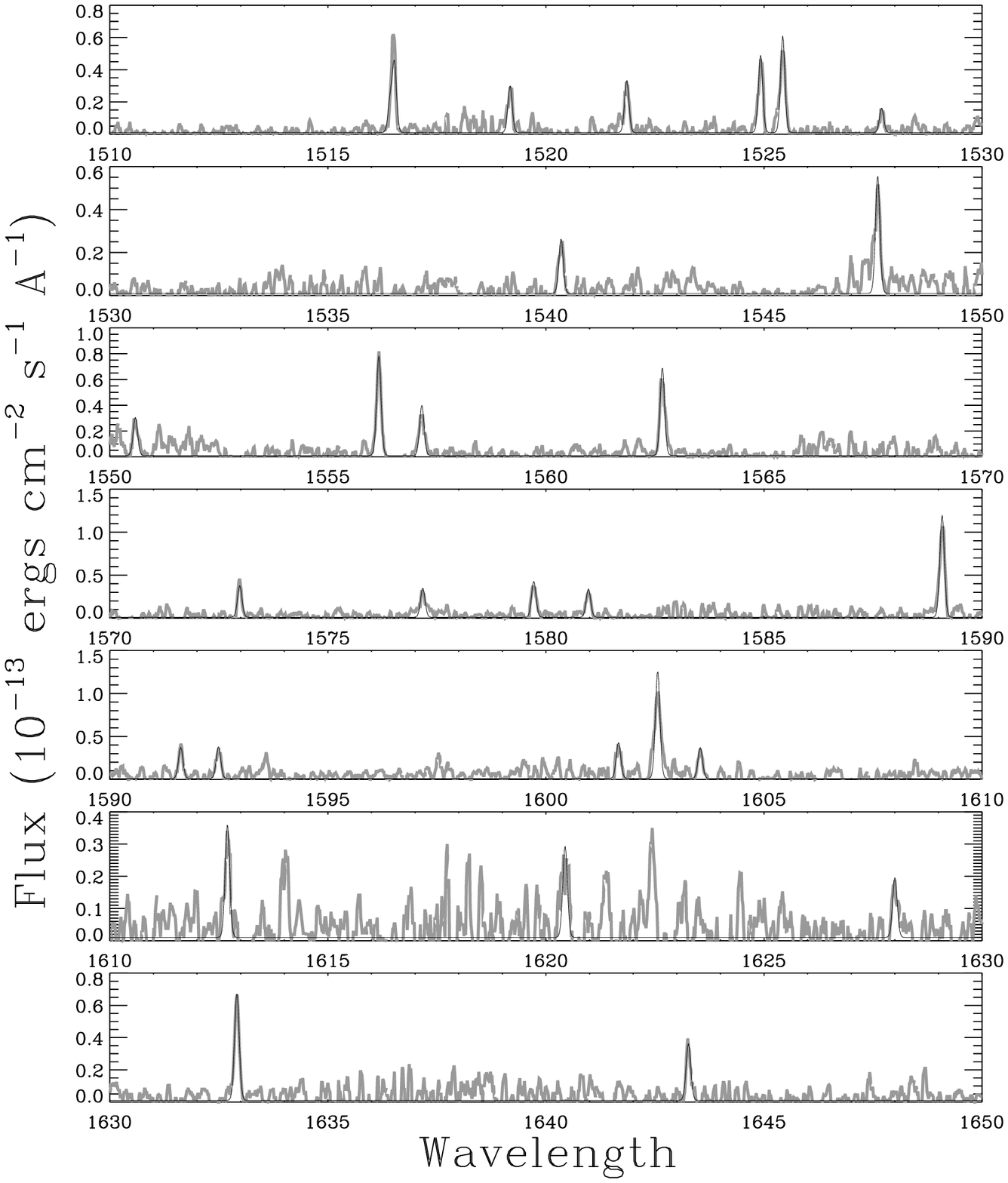}{6.5in}{0}{80}{80}{-340}{10}
\caption{(continued)}
\end{figure}

\clearpage

\begin{figure}
\plotfiddle{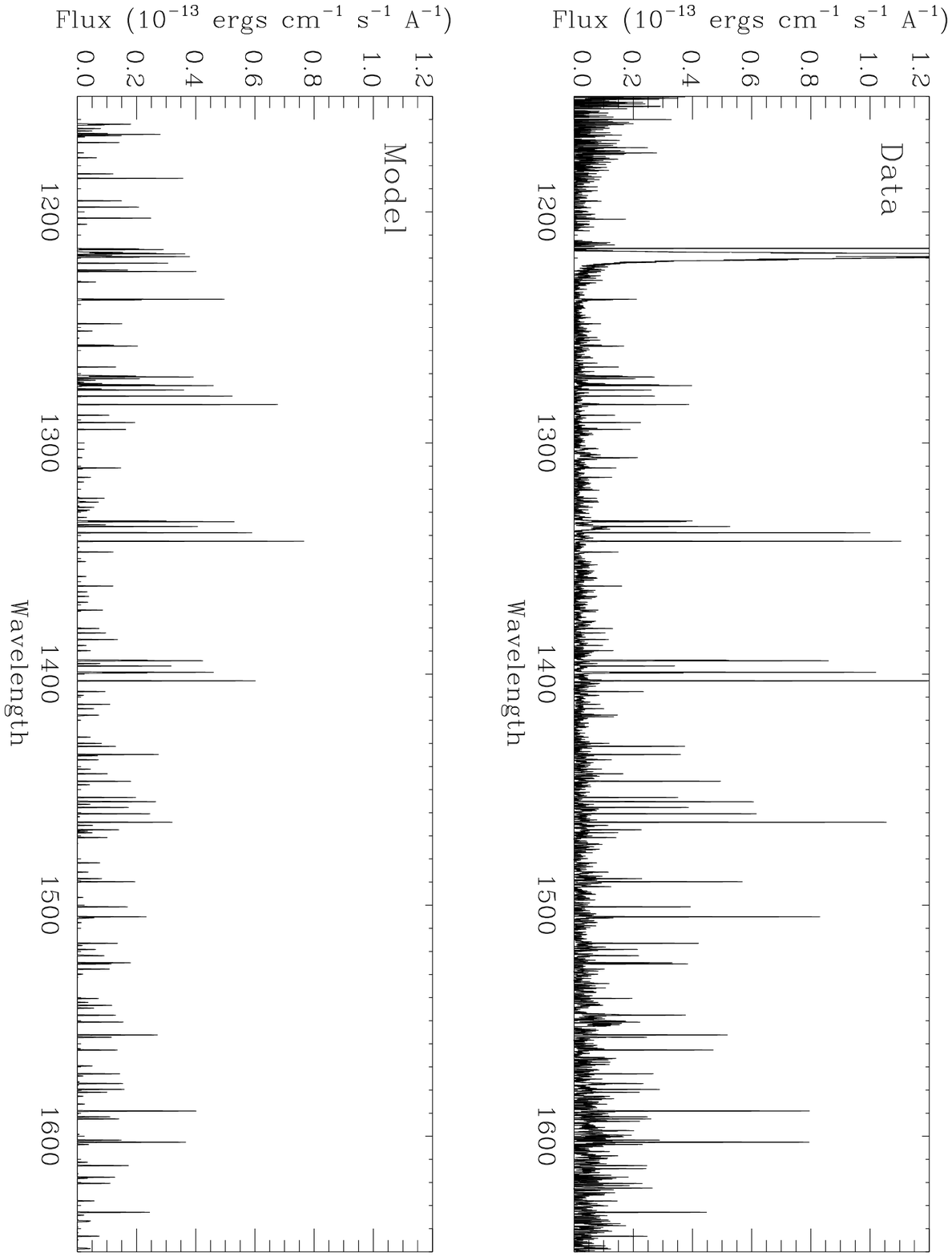}{3.5in}{90}{75}{75}{290}{0}
\caption{The top panel is a smoothed representation of the full E140M
  spectrum of Mira~B, showing numerous narrow H$_{2}$ lines produced by
  Ly$\alpha$ fluorescence.  The bottom panel is a fit to this spectrum
  assuming that the line ratios within the 13 identified fluorescence
  sequences are consistent with the H$_{2}$ transition probabilities.  The
  fit is very poor and the discrepancies are wavelength dependent,
  presumably due to an opacity effect that results in the observed fluxes
  of low wavelength H$_{2}$ lines being lower than they should relative
  to higher wavelength lines.}
\end{figure}

\clearpage

\begin{figure}
\plotfiddle{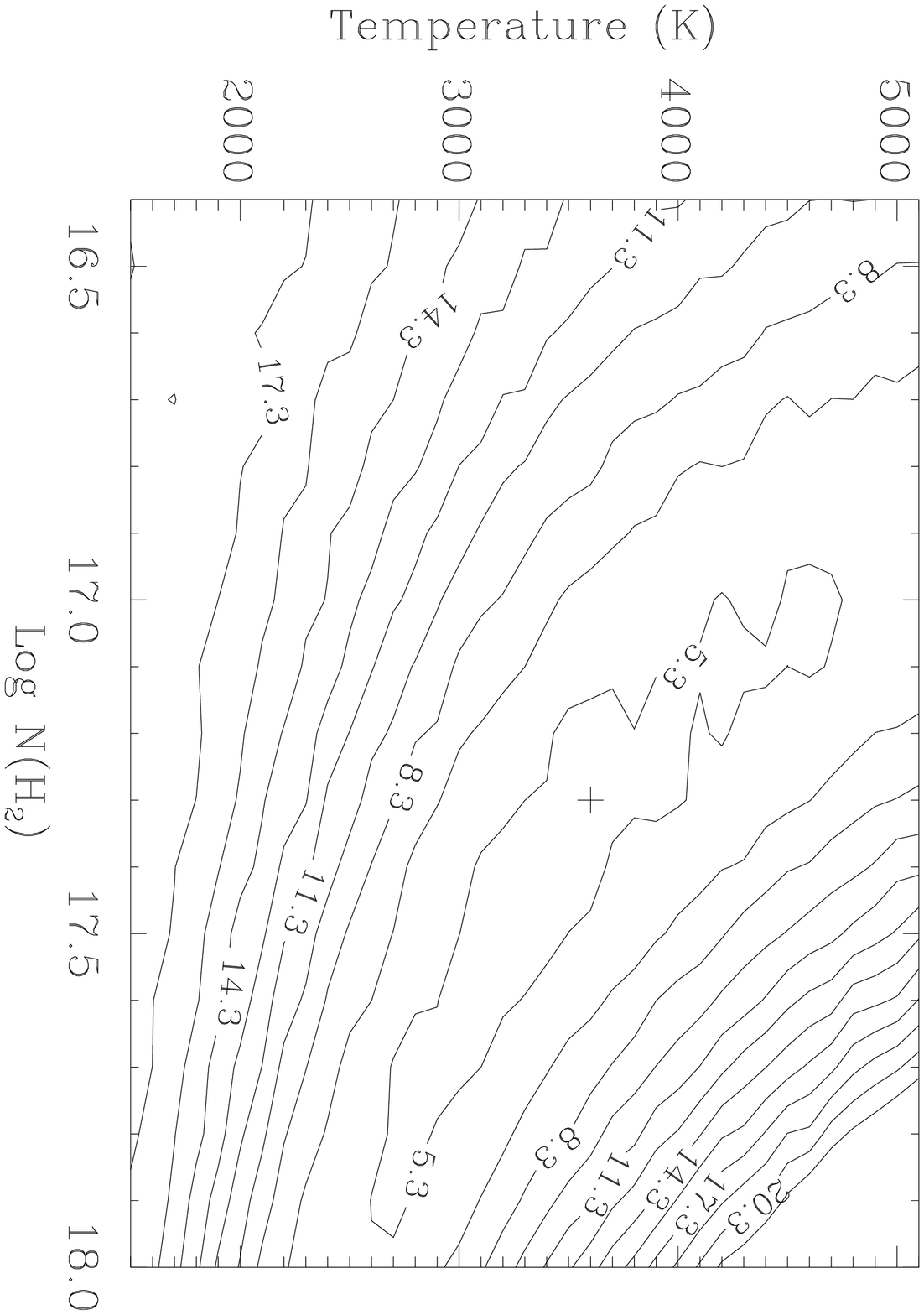}{3.5in}{90}{75}{75}{290}{0}
\caption{Reduced chi-squared contours based on the Monte Carlo radiative
  transfer calculation described in the text, which is meant to reproduce
  the observed H$_{2}$ line ratios.  The best fit to the data
  is obtained when assuming an H$_{2}$ temperature and column density of
  $T=3600$~K and $\log {\rm N(H_{2})}=17.3$, respectively (the plus sign).}
\end{figure}

\clearpage

\begin{figure}
\plotfiddle{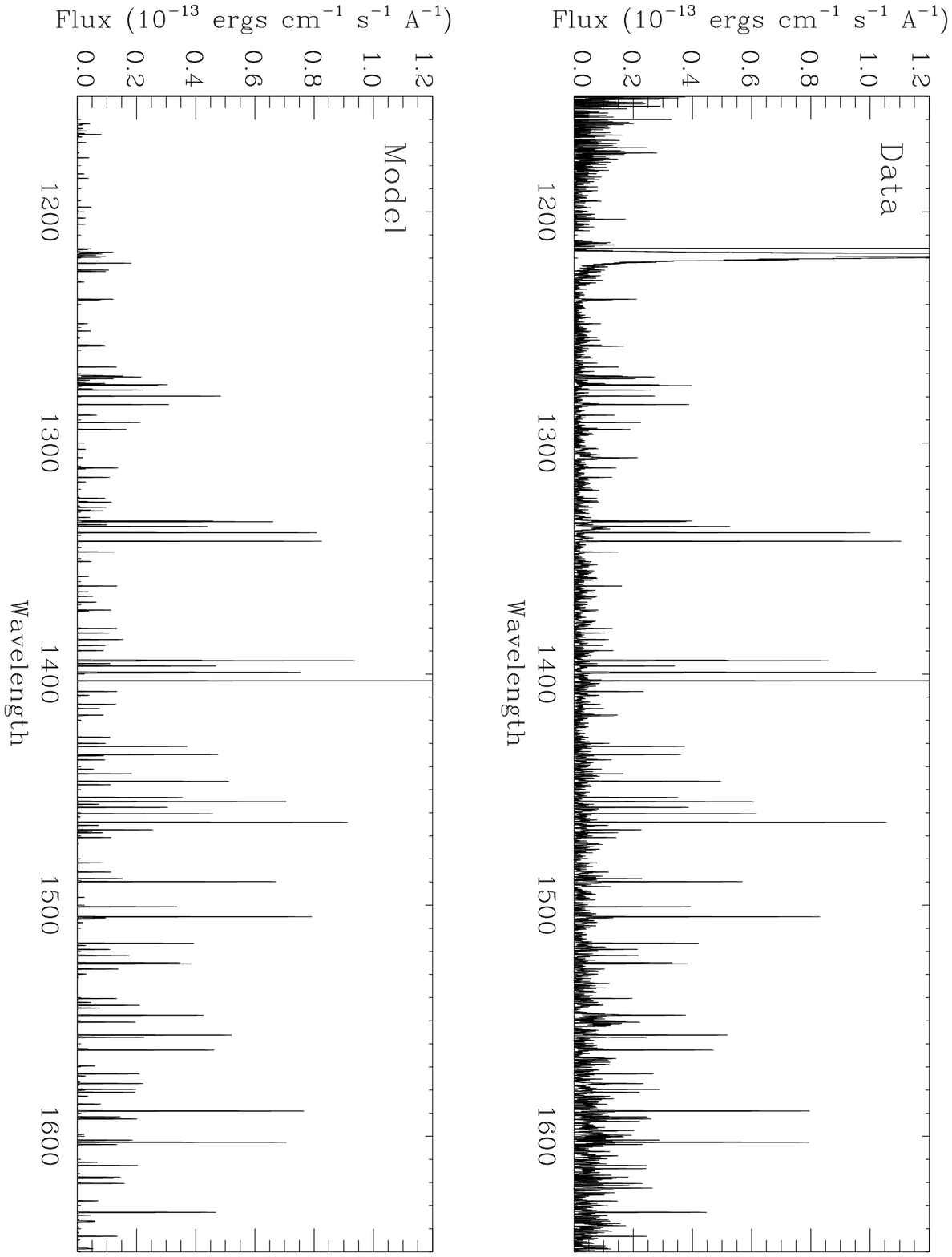}{3.5in}{90}{75}{75}{290}{0}
\caption{A comparison of the observed spectrum (top panel) and a model
  spectrum (bottom panel) fitted to the data as in Fig.~2, but with the
  consideration of opacity effects, assuming an H$_{2}$ fluorescence layer
  with a temperature and column density of $T=3600$~K and
  $\log {\rm N(H_{2})}=17.3$, respectively (the best fit from Fig.~3).
  Consideration of the opacity effects greatly improves the quality of the
  fit compared to that in Fig.~2.}
\end{figure}

\clearpage

\begin{figure}
\plotfiddle{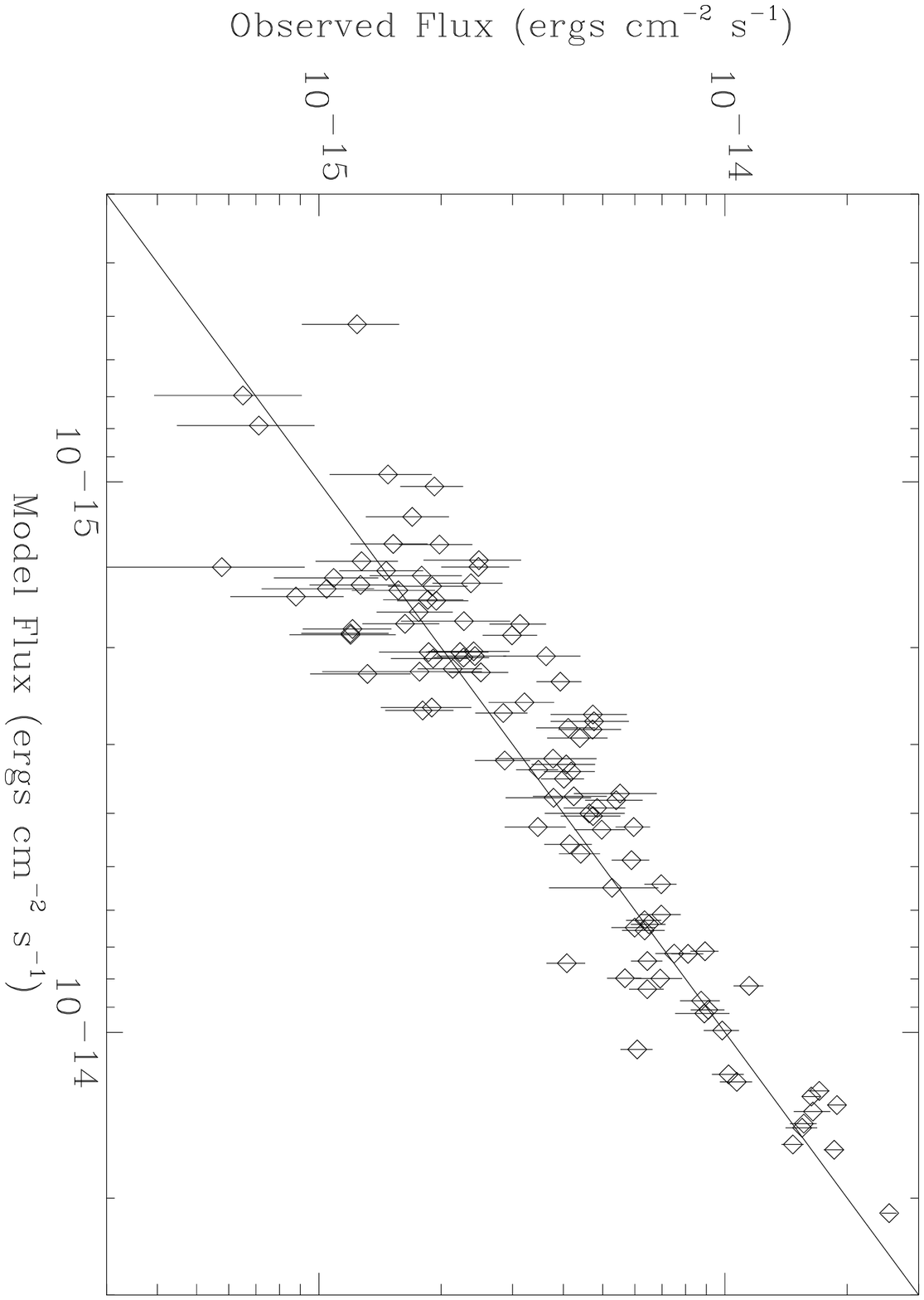}{3.5in}{90}{75}{75}{295}{0}
\caption{A plot of measured H$_{2}$ line fluxes versus the predicted
  fluxes of the model illustrated in Fig.~4.  These observed and model
  fluxes are all listed in Table~1.}
\end{figure}

\clearpage

\begin{figure}
\plotfiddle{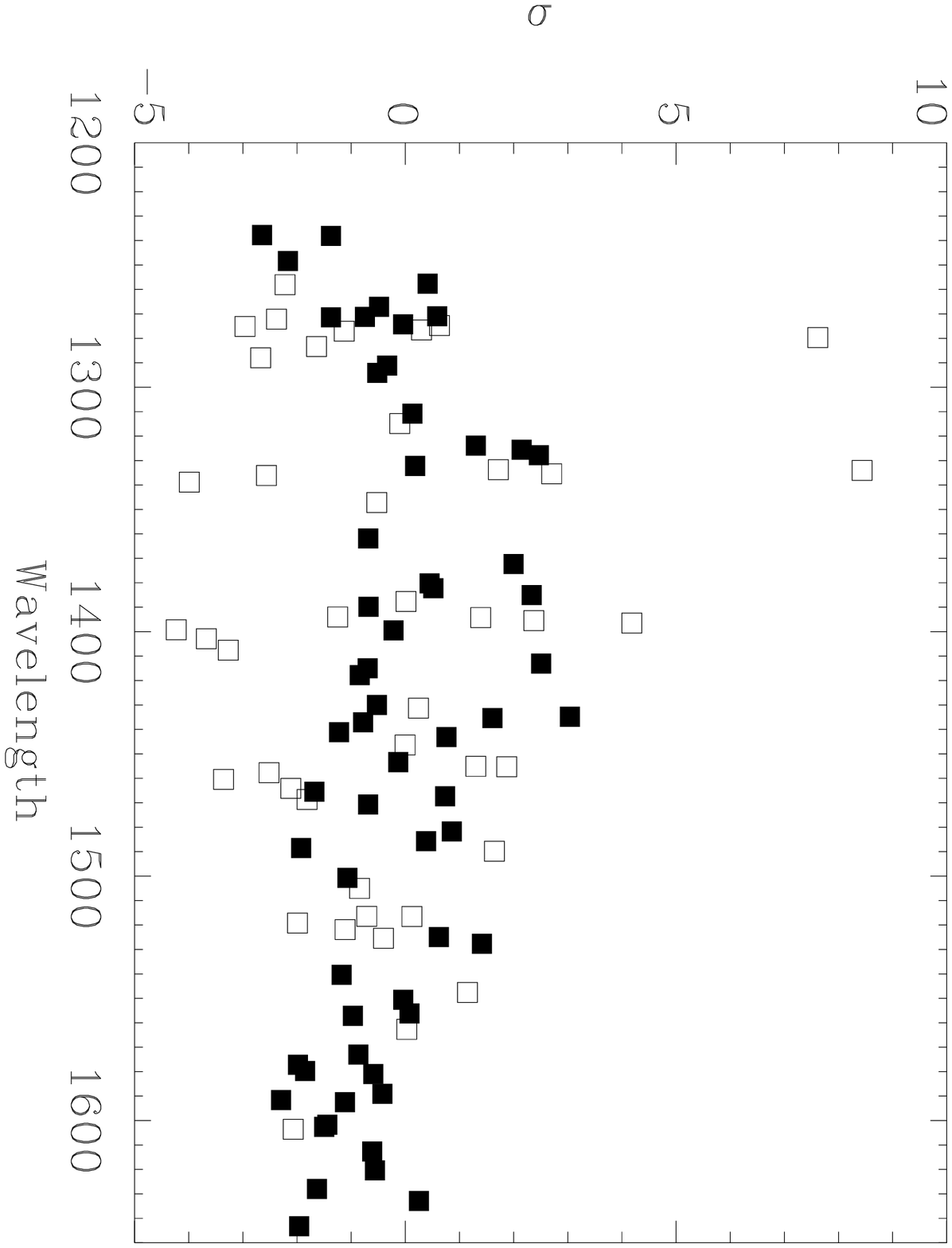}{3.5in}{90}{75}{75}{280}{0}
\caption{The discrepancies between the measured and modeled line fluxes
  from Fig.~5, plotted in standard deviation units as a function of
  wavelength.  Lines in fluorescence sequences that are excited from low
  energy levels ($E_{low} < 10,000$ cm$^{-1}$) are shown as open boxes and
  lines of fluorescence sequences excited from high energy levels
  ($E_{low} > 10,000$ cm$^{-1}$) are shown as filled boxes (see Table~2).
  The low energy lines are systematically more discrepant than the high
  energy lines.}
\end{figure}

\clearpage

\begin{figure}
\plotfiddle{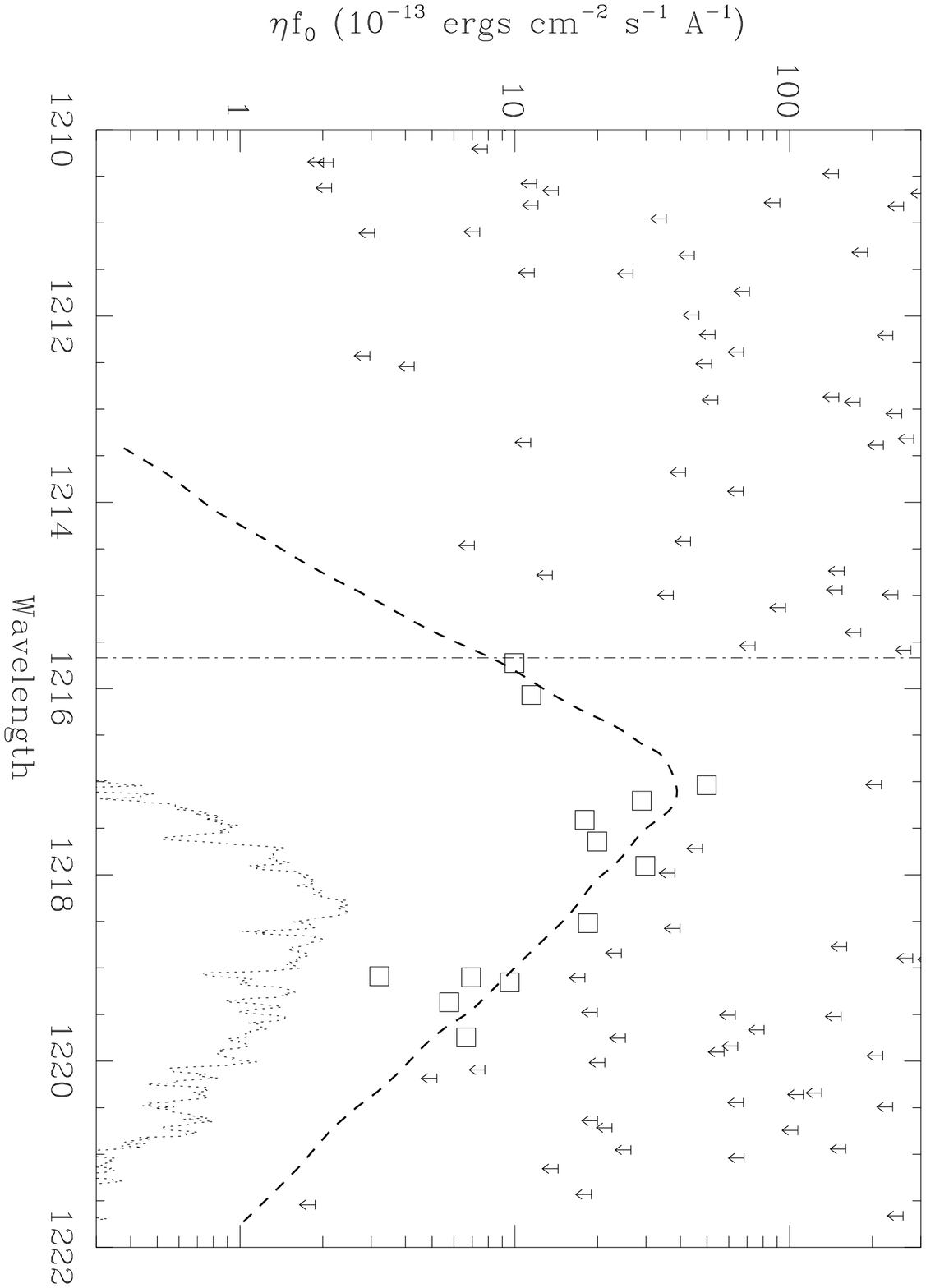}{3.5in}{90}{75}{75}{290}{0}
\caption{The background Ly$\alpha$ fluxes seen by the fluoresced H$_{2}$
  are determined using the total fluoresced fluxes of the 13 detected
  fluorescence sequences, and these background fluxes are plotted as a
  function of wavelength (boxes).  Extrapolation between these data points
  yields an estimate of the Ly$\alpha$ profile seen by the H$_{2}$ (dashed
  line).  In this analysis, we assume a temperature and column density for
  the H$_{2}$ layer of $T=3600$~K and $\log {\rm N(H_{2})}=17.3$,
  respectively (the best fit from Fig.~3).  The observed Ly$\alpha$ profile
  (slightly smoothed) is shown as a dotted line.  Upper limits for
  background Ly$\alpha$ fluxes are also shown based on undetected H$_{2}$
  fluorescence sequences, which are consistent with the detections (boxes).
  The vertical dot-dashed line is the rest frame of Mira~B.}
\end{figure}

\clearpage

\begin{figure}
\plotfiddle{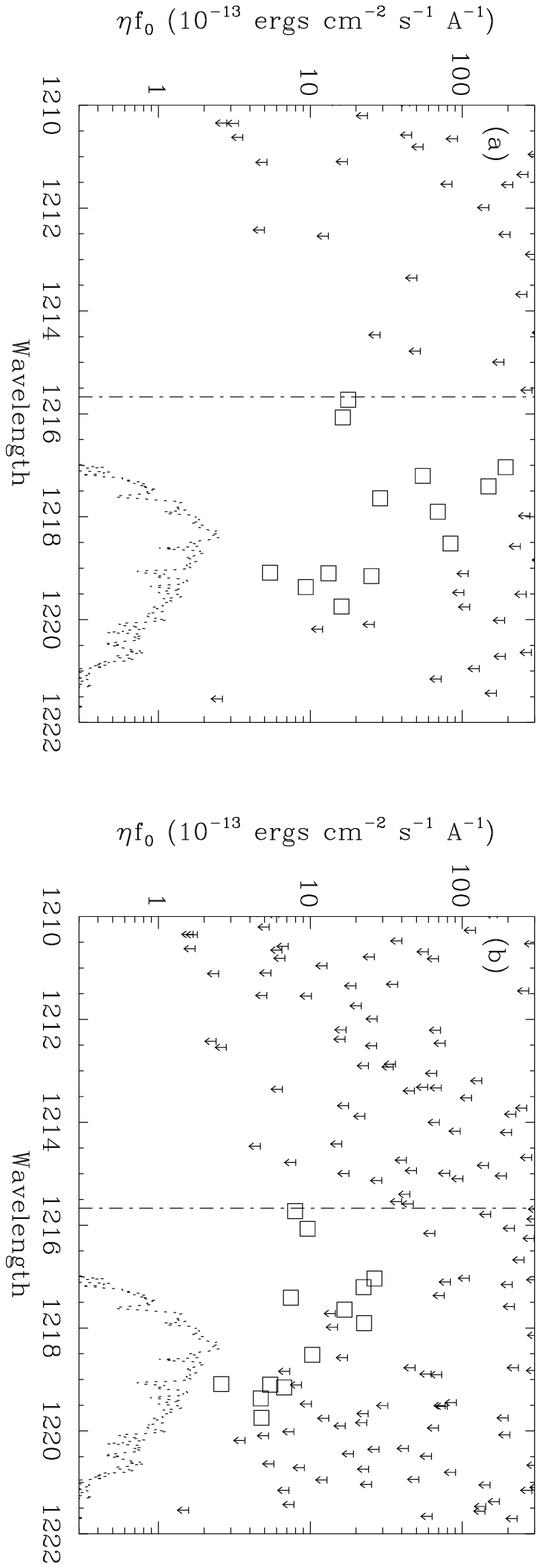}{2.0in}{90}{75}{75}{290}{-150}
\caption{Two figures analogous to Fig.~7, but assuming different H$_{2}$
  temperatures: (a) $T=2600$~K, and (b) $T=4600$~K.  The lower temperature
  clearly leads to greater scatter in the derived background Ly$\alpha$
  fluxes (boxes) and the higher temperature results in some discrepant
  upper limits.  This provides support for the intermediate $T=3600$~K
  temperature derived in Fig.~3 and assumed in Fig.~7.}
\end{figure}

\clearpage

\begin{figure}
\plotfiddle{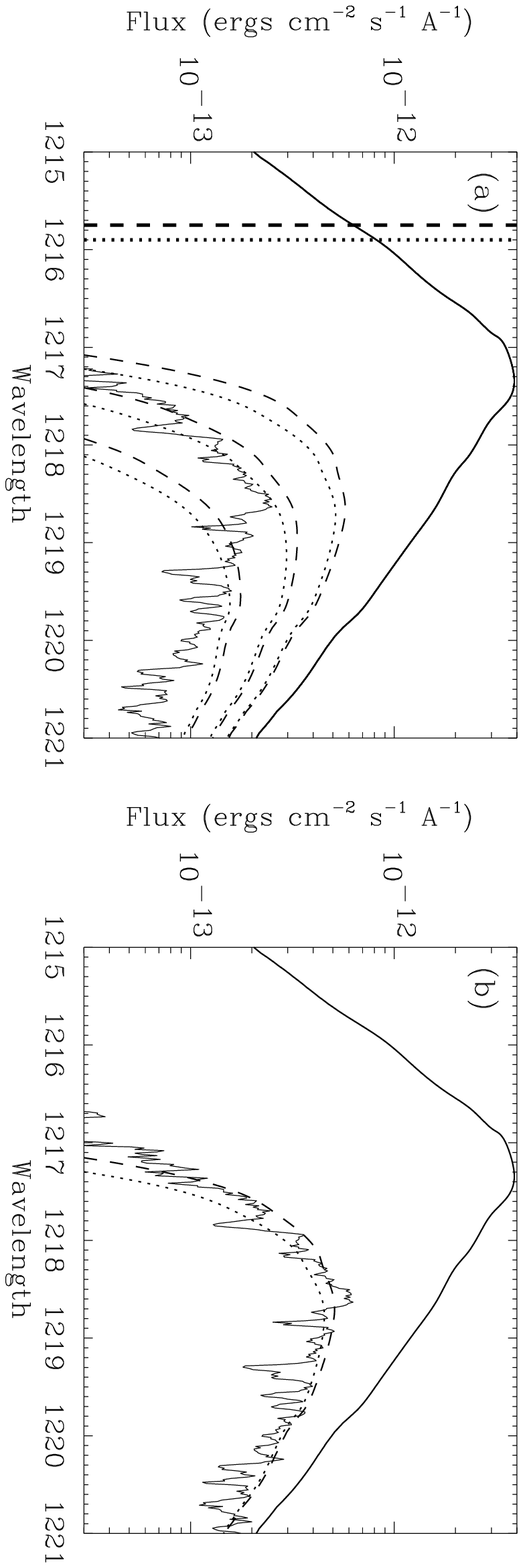}{2.0in}{90}{75}{75}{290}{-150}
\caption{(a) The lower and upper solid lines are the observed Ly$\alpha$
  profile and the Ly$\alpha$ profile seen by the fluoresced H$_{2}$,
  respectively (from Fig.~7).  The estimated center of ISM H~I absorption
  is shown as a vertical dashed line, and the stellar rest frame is
  indicated by the vertical dotted line.  Interstellar and circumstellar
  H~I absorption profiles (dashed and dotted lines, respectively) are
  computed assuming column densities of $\log {\rm N(H~I)}=20.3$, 20.5, and
  20.7.  None fit the data well.  (b) Interstellar and circumstellar H~I
  absorption for $\log {\rm N(H~I)}=20.35$, compared with the observed
  Ly$\alpha$ profile multiplied by a factor of 2.5, which yields a
  reasonable fit.}
\end{figure}

\clearpage

\begin{figure}
\plotfiddle{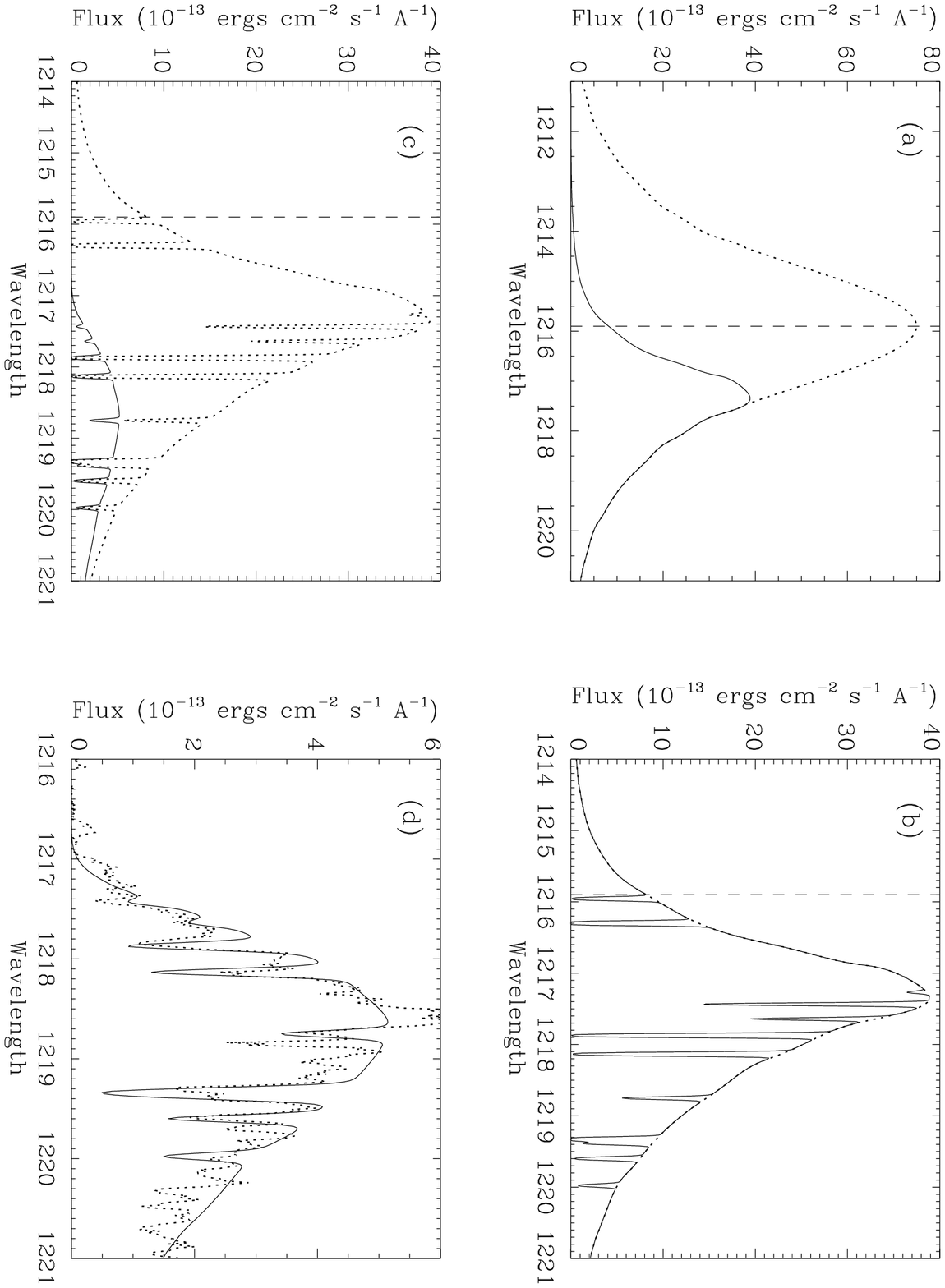}{3.5in}{90}{75}{75}{290}{0}
\caption{The evolution of the H~I Ly$\alpha$ profile during its journey
  from Mira~B to the Earth. Each panel shows a step in the evolution of the
  profile.  The vertical dashed line is the rest frame of  Mira~B.  (a) The
  profile emitted by Mira~B's accretion disk (dotted line), and the profile
  after absorption from Mira~B's wind (solid line).  (b) The profile before
  (dotted) and after (solid) H$_{2}$ absorption.  (c) The profile before
  (dotted) and after (solid) interstellar (or circumstellar) H~I absorption.
  (d) The profile smoothed by the STIS E140M line spread function (solid
  line), compared with a smoothed representation of the observed profile
  multiplied by a factor of 2.5 (dotted line).}
\end{figure}

\clearpage

\begin{figure}
\plotfiddle{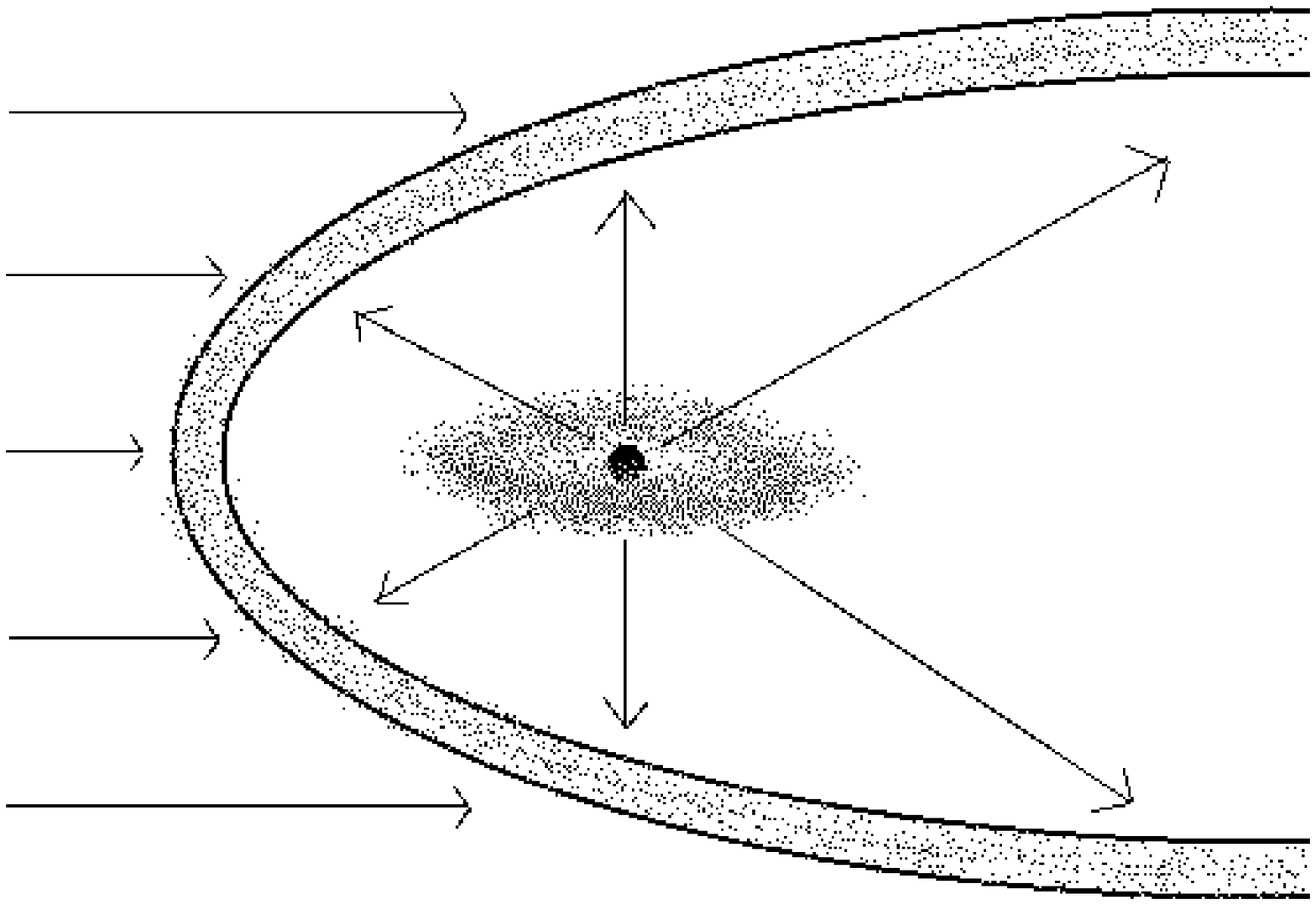}{3.5in}{0}{80}{80}{-250}{-150}
\caption{A schematic diagram of Mira~B's radial wind interacting with 
  that of Mira~A entering from the left.  Mira~B is shown surrounded by
  an accretion disk.  One interpretation of the H$_{2}$ emission seen by
  HST/STIS is that it arises in the wind interaction region.}
\end{figure}

\clearpage

\begin{figure}
\plotfiddle{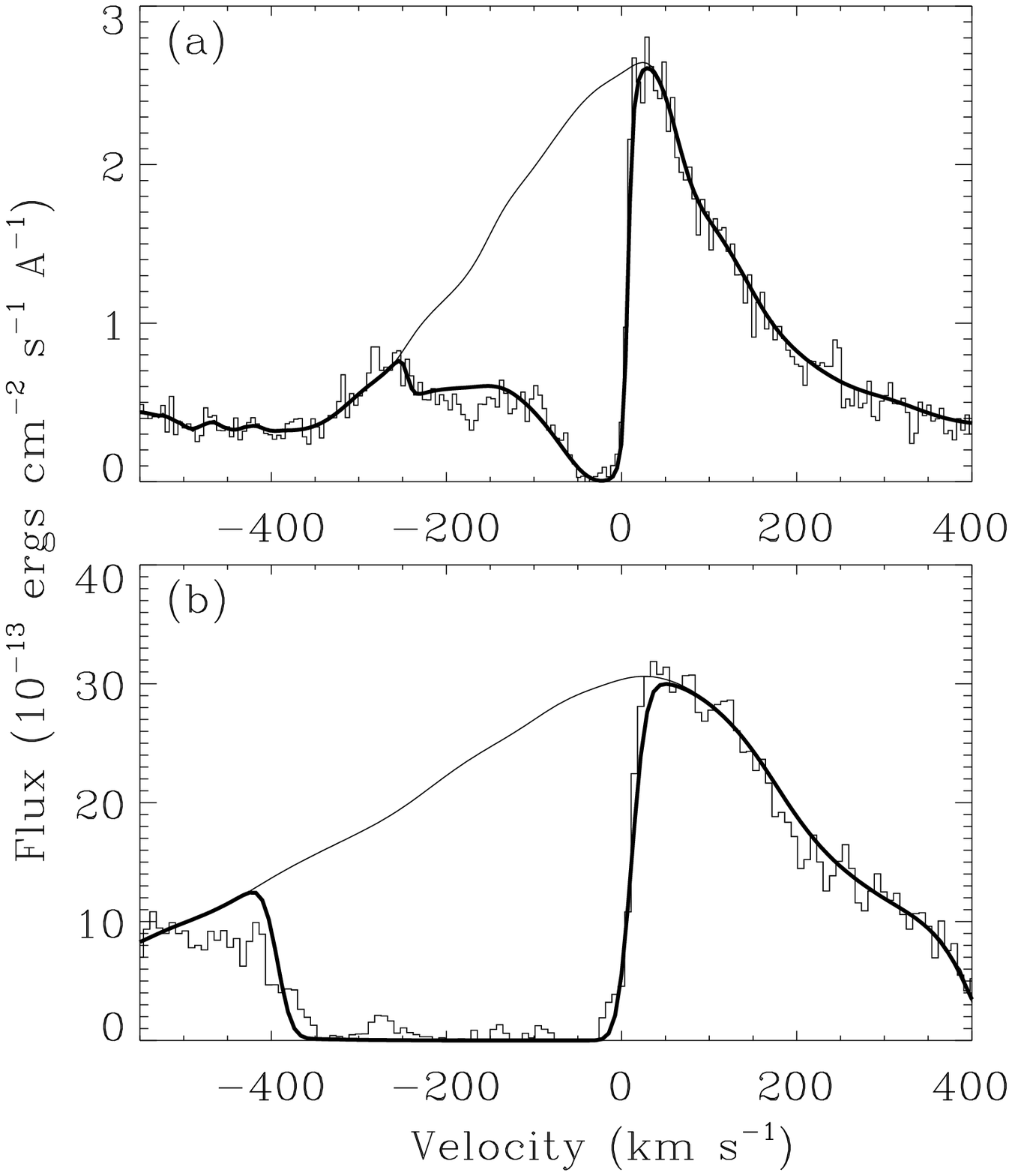}{5.0in}{0}{80}{80}{-270}{10}
\caption{(a) The Mg~II k line profile observed by HST/STIS, plotted on a
  velocity scale centered on the stellar rest frame.  The intrinsic line
  profile above the wind absorption feature is estimated (thin solid line),
  and the wind absorption is modeled as described in the text to determine
  the mass loss rate ($\dot{M}$) and termination velocity ($V_{\infty}$)
  that yield the best fit to the data, which is shown as a thick solid line.
  For this fit, $\dot{M}=5\times 10^{-13}$ M$_{\odot}$ yr$^{-1}$ and
  $V_{\infty}=250$ km~s$^{-1}$.  (b) A typical Mg~II k line profile
  observed by IUE (LWP 29795), with the wind absorption analyzed in the same
  fashion as in (a).  For this fit,
  $\dot{M}=1\times 10^{-11}$ M$_{\odot}$ yr$^{-1}$
  and $V_{\infty}=400$ km~s$^{-1}$.}
\end{figure}

\clearpage

\begin{figure}
\plotfiddle{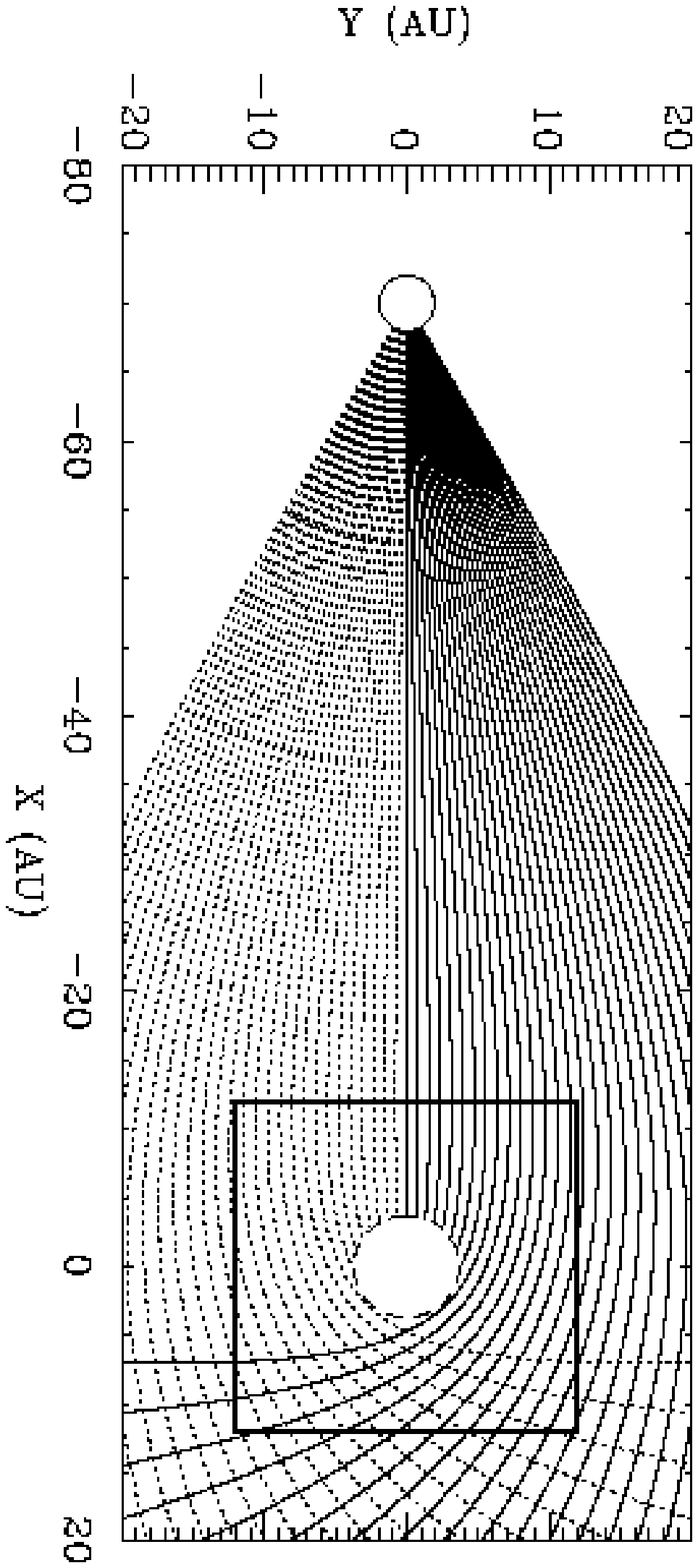}{3.5in}{90}{75}{75}{295}{-20}
\caption{Particle trajectories for Mira~A's wind under the influence of
  Mira~B's gravitational field.  Mira~A is the 2~AU radius circle on the
  left.  The trajectories are truncated within 3.7~AU of Mira~B (dashed
  circle) since that is roughly the distance where the ram pressures of
  the winds of Mira~A and Mira~B balance, at least in the direction of
  Mira~A.  The large square is the projected size of the STIS aperture
  used in our Mira~B observations.}
\end{figure}

\clearpage

\begin{figure}
\plotfiddle{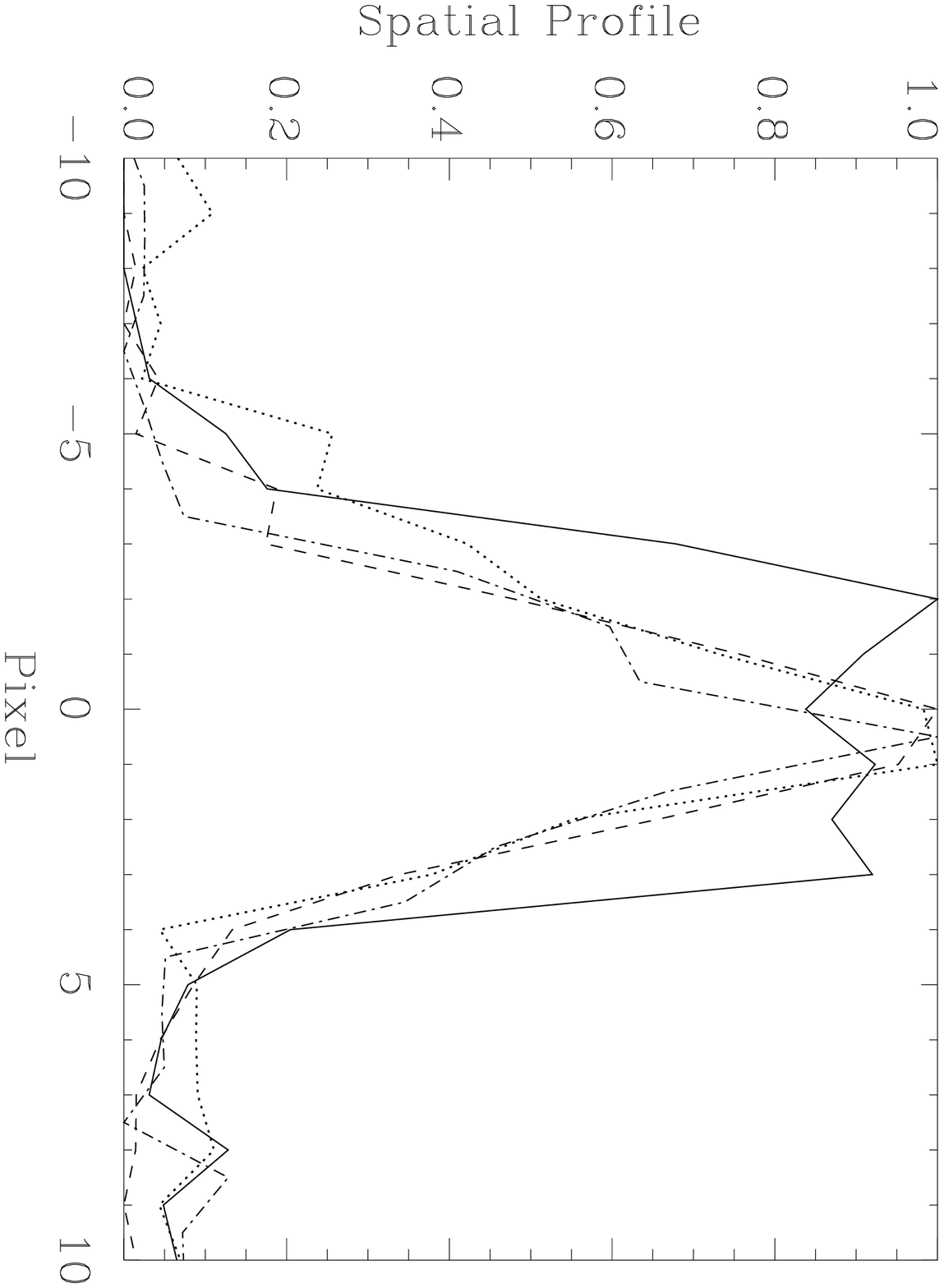}{3.5in}{90}{75}{75}{295}{0}
\caption{Spatial profiles of several spectral features within our E140M
  spectrum of Mira~B:  Geocoronal Ly$\alpha$ (solid line), stellar
  Ly$\alpha$ (dotted line), H$_{2}$ 0-4 P(3) (dashed line), and a blend of
  the H$_{2}$ 0-5 R(0) and 0-5 R(1) lines (dot-dashed lines).  The
  geocoronal Ly$\alpha$ line fills the aperture and is therefore broader
  than the other profiles and non-Gaussian.  We see no evidence that the
  other emission lines are spatially resolved.}
\end{figure}

\clearpage

\begin{figure}
\plotfiddle{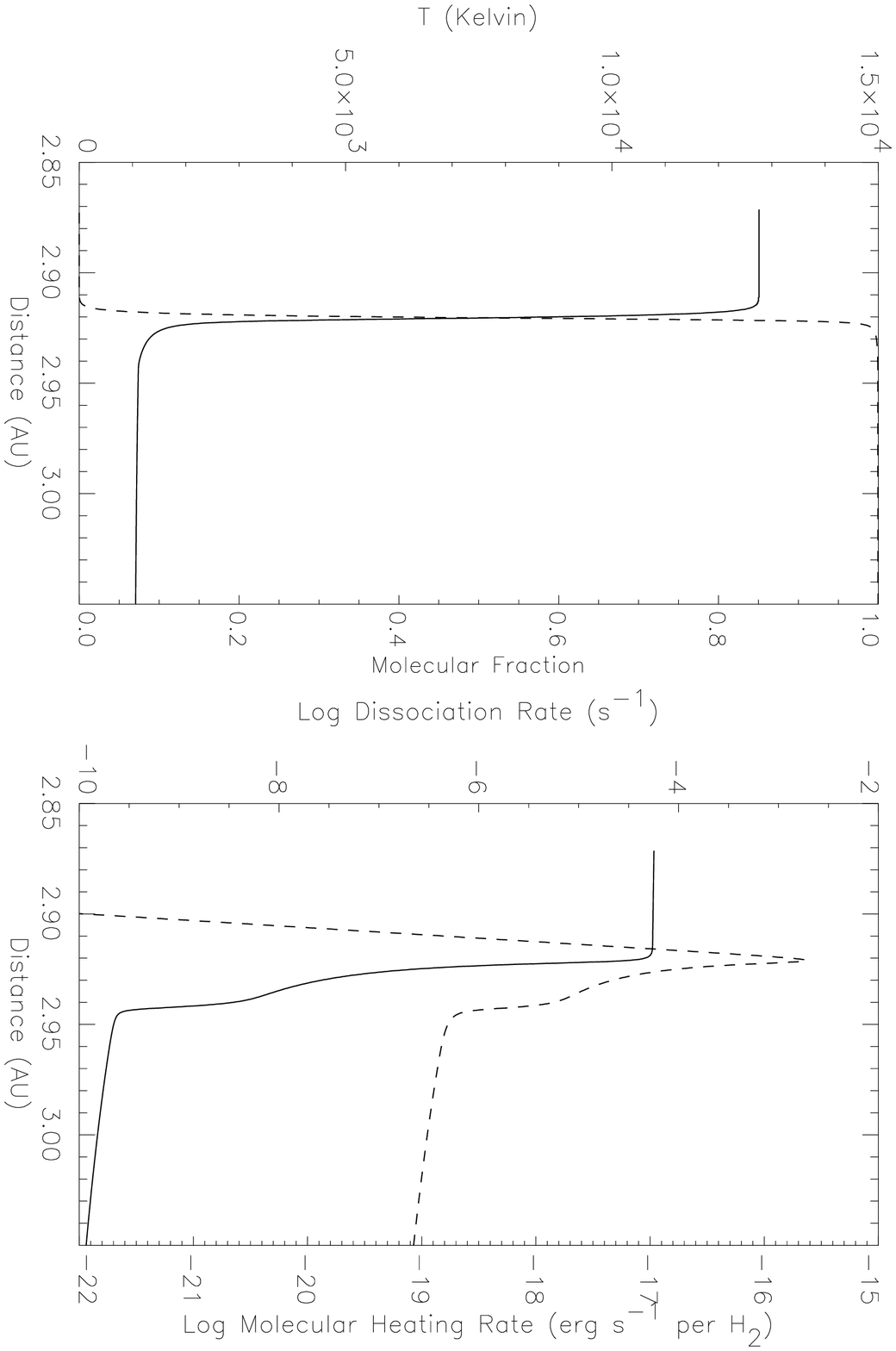}{3.5in}{90}{65}{65}{240}{-20}
\caption{A model of a photodissociation front.  The left panel
  is the kinetic temperature (solid line) and molecular fraction for
  hydrogen (dashed line) as a function of distance from Mira~B, and the
  right panel is the H$_{2}$ dissociation rate (solid line) and heating
  rate due to fluorescence excitation and dissociation (dashed line).}
\end{figure}

\clearpage

\begin{figure}
\plotfiddle{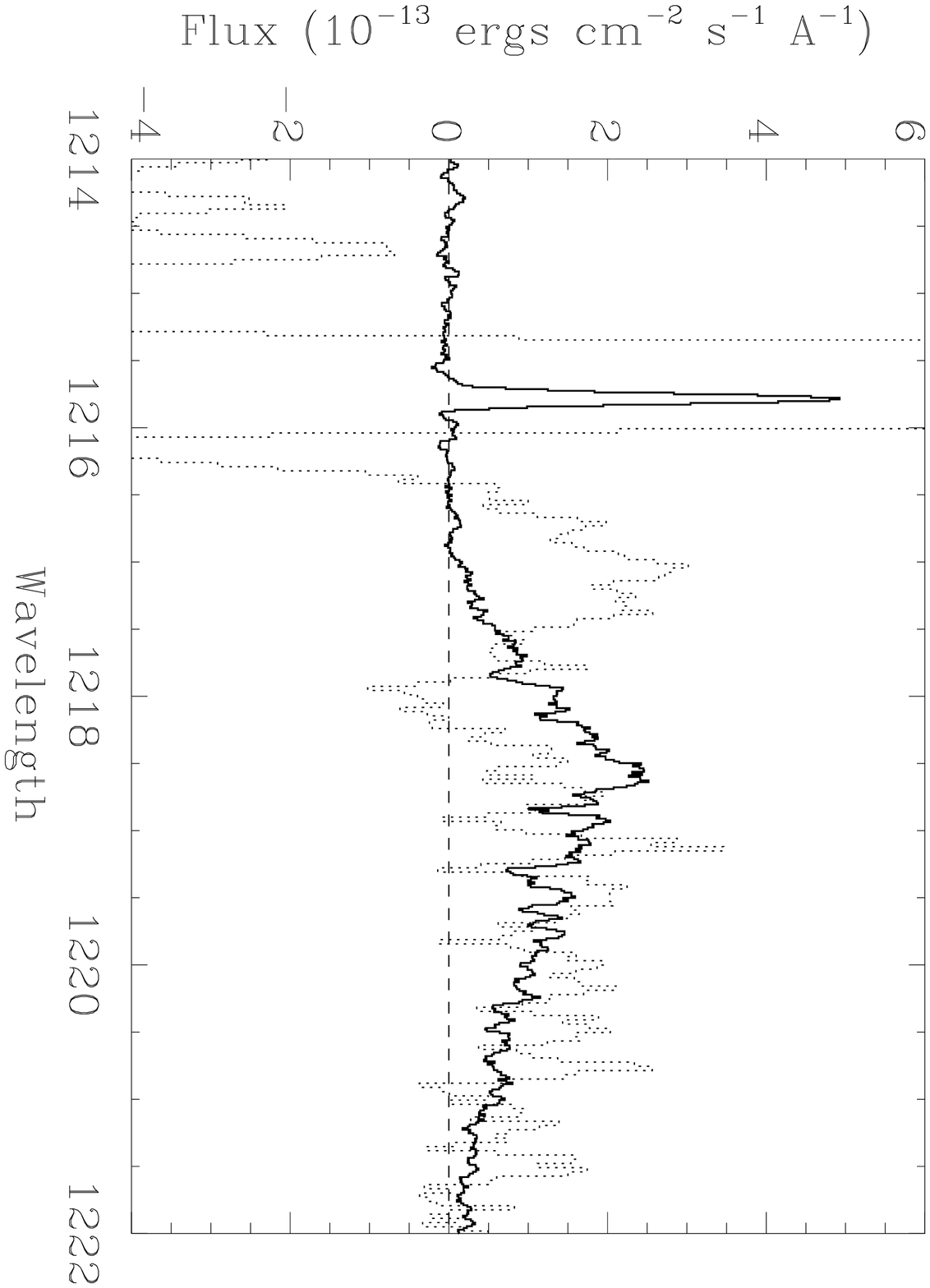}{3.5in}{90}{75}{75}{295}{0}
\caption{Comparison between the H~I Ly$\alpha$ profile observed by STIS
  (solid line) and a profile observed by IUE (dotted line), based on a
  coaddition of three separate IUE spectra.  Geocoronal emission is seen in
  both spectra near 1215.8~\AA.  Both spectra have been smoothed.  The
  fluxes of the IUE spectrum are not trustworthy due to background
  subtraction difficulties, but the main purpose of the comparison is
  simply to show that the IUE Ly$\alpha$ fluxes are clearly not an order of
  magnitude higher than the STIS fluxes, in contrast to the flux behavior
  of the UV continuum and all other non-H$_{2}$ emission lines.}
\end{figure}

\end{document}